\documentclass{iopart}

 \expandafter\let\csname equation*\endcsname\relax
  \expandafter\let\csname endequation*\endcsname\relax
\usepackage{amsmath}

\usepackage{amssymb}

\usepackage{graphicx}

\usepackage{cite}

\usepackage{color} 
\usepackage{amsmath,amssymb,graphics,graphicx} 

\usepackage[bookmarks,colorlinks,breaklinks]{hyperref}  
\usepackage{setspace}
\usepackage{multirow}
\usepackage{color}
\usepackage{rotating}
\usepackage{verbatim}

\usepackage[labelfont=bf,labelsep=period,justification=raggedright]{caption}

\makeatletter
\renewcommand{\@biblabel}[1]{\quad#1.}
\makeatother

\date{}

\begin{document}
\title{Inter-dependent Tissue Growth and Turing Patterning in a Model for Long Bone Development}

\author{Simon Tanaka$^{1}$ and Dagmar Iber$^{1,2\ast}$}

\address{1 Department for Biosystems Science and Engineering, ETH Zurich, Mattenstrasse 26, Basel, Switzerland,  +41 61 387 32 10 (phone), +41 61 387 31 94 (fax)\\ 
2 Swiss Institute of Bioinformatics, Basel, Switzerland}
\ead{dagmar.iber@bsse.ethz.ch}



\begin{abstract}
The development of long bones requires a sophisticated spatial organization of cellular signaling, proliferation,
and differentiation programs. How such spatial organization emerges on the growing long bone domain is still unresolved.
Based on the reported biochemical interactions we developed a regulatory model for the core signaling factors IHH, PTCH1, and PTHrP and included two cell types, proliferating/resting chondrocytes and (pre-)hypertrophic chondrocytes. We show that the reported IHH-PTCH1 interaction gives rise to a Schnakenberg-type Turing kinetics, and that inclusion of PTHrP is important to achieve robust patterning when coupling patterning and tissue dynamics.
The model reproduces relevant spatiotemporal gene expression patterns, as well as a number of relevant mutant phenotypes.
In summary, we propose that a ligand-receptor based Turing mechanism may control the emergence of patterns during long bone development, with PTHrP as an important mediator to confer patterning robustness when the sensitive Turing system is coupled to the dynamics of a growing and differentiating tissue. We have previously shown that ligand-receptor based Turing mechanisms can also result from BMP-receptor and GDNF-receptor interactions, and that these reproduce the wildtype and mutant patterns during digit formation in limbs and branching morphogenesis in lung and kidneys. Receptor-ligand interactions may thus constitute a general mechanism to generate Turing patterns in nature.
\end{abstract}



\submitto{Physical Biology}

\maketitle

\section{Introduction}

Long bones develop by endochondral ossification \cite{Kronenberg:2003p47895, Provot:2005fj, Wuelling:2011dw}.
As part of the process, mesenchymal stem cells (which are the multipotent precursors of cells involved in bone formation) aggregate in condensations and differentiate into chondrocytes (Fig. \ref{fig:Fig2}).
As the bone develops, different types of chondrocytes emerge in different parts of the condensation and express different sets of genes.
Most prominently proliferating and resting chondrocytes are found at the ends of the bone primordium within well-defined domains, while hypertrophic chondrocytes emerge in the center of the domain.
How this spatial pattern emerges from the regulatory interactions is so far unresolved.  

The patterning process occurs on a growing domain.
Growth of the bone domain is a consequence of both proliferation and differentiation into larger cells \cite{Cooper2013}.
Cell differentiation into hypertrophic chondrocytes, and, at later stages, apoptosis of hypertrophic chondrocytes and replacement by invading osteoblasts all start in the center of the domain \cite{Wuelling:2010ev,Wuelling:2011dw}. 

The core regulatory network has been defined to comprise Parathyroid hormone-related protein (PTHrP), Indian Hedgehog (IHH), and its receptor PTCH1 \cite{Lanske:1996vc,Vortkamp:1996wz,Kronenberg:2003p47895}.
In what follows, we use the following convention: 
protein names are written all-capitalized (e.g. IHH for the protein Indian Hedgehog), and their gene names are italicized (e.g. \textit{Ihh}).
Furthermore, the expression of a gene is defined as the production rate of the protein (e.g. \textit{Ihh} expression is the production rate of IHH).
IHH signaling is the event of binding of a IHH protein to its receptor (and thus triggering downstream regulatory processes) and therefore corresponds to the concentration of the ligand-receptor complex.
The abbreviated gene and protein names are summarized in Table \ref{tab:teminology}.

\begin{figure}[h!]
\begin{centering}
\includegraphics[width=0.8\columnwidth]{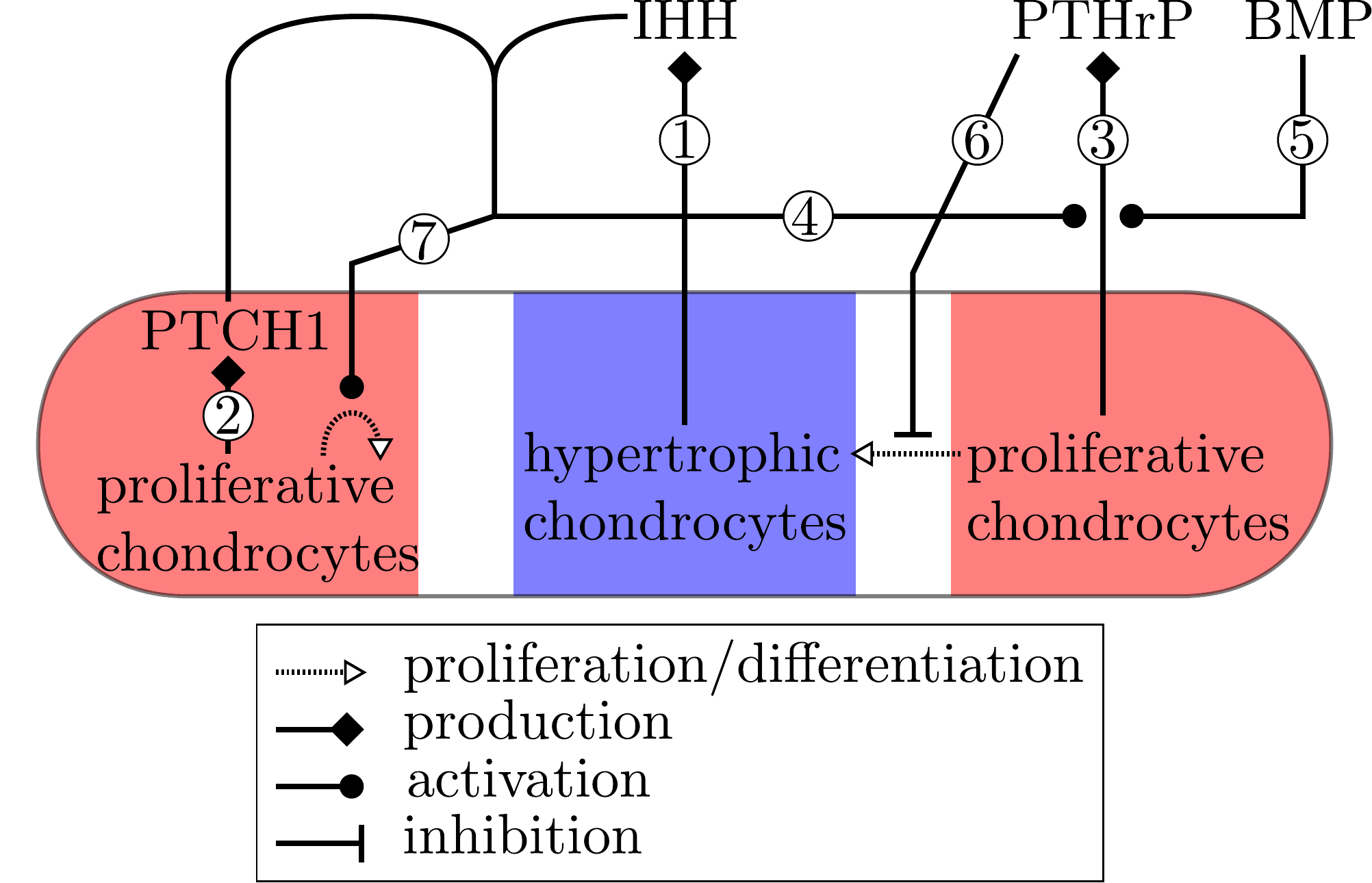}
\par\end{centering}
\caption{}
\label{fig:Fig1}
\end{figure}

The gene \textit{Ihh} is expressed mainly by hypertrophic chondrocytes in the center of the domain (Figure \ref{fig:Fig1}, arrow 1; abbreviated as A1 going forward), but  expression of \textit{Ihh} can be detected already around embryonic day (E)10.5 (days post coitum) in the forelimb before hypertrophic chondrocytes emerge.
The gene \textit{Ptch1}, which encodes the IHH receptor PTCH1, is expressed mainly by proliferating and resting chondrocytes \cite{Hilton:2005cu} (Figure \ref{fig:Fig1}, A2). The gene \textit{Pthrp} is expressed by resting periarticular chondrocytes that reside at the ends of the domain \cite{Karp:2000wm} (Figure \ref{fig:Fig1}, A3). The expression of  \textit{Pthrp}  is controlled by IHH (Figure \ref{fig:Fig1}, A4) \cite{Vortkamp:1996wz, Kobayashi:2005ge} and BMP signaling \cite{Pateder:2001dd,Pateder:2000kb,Zhang:2003gh,Zou:1997wm} Figure \ref{fig:Fig1}, A5).
PTHrP is a diffusible, extracellular protein that  increases the pool of mitotically active (i.e. proliferating) chondrocytes by preventing their differentiation into hypertrophic chondrocytes (Figure \ref{fig:Fig1}, A6) \cite{Karaplis:1994vh, Weir:1996wf}, but, unlike Hedgehog signaling, PTHrP does not enhance their proliferation rate (Figure \ref{fig:Fig1}, A7 \cite{Karp:2000wm}. 

While the core regulatory network has been resolved, it has remained unclear how the patterns and the spatio-temporal control of the process emerge from these interactions. A number of mathematical models have been developed to explain the distribution of the signaling proteins IHH and PTHrP and their impact on bone growth and development \cite{Bougherara:2010p47089, vanDonkelaar:2007p47860, Brouwers:2006p47851, Isaksson:2008p47849, GarzonAlvarado:2009p47095,GarzonAlvarado:2010p47081}.
Garzon-Alvarado and colleagues suggest a Schnakenberg-type Turing patterning mechanism based on regulatory interactions between IHH and PTHrP  \cite{GarzonAlvarado:2009p47095}. In particular, they postulate that the rate of PTHrP production and  IHH removal are both proportional to the concentration of PTHrP squared times the IHH concentration ($[\text{PTHrP}]^2 [\text{Ihh}]$). While IHH signaling indeed enhances \textit{Pthrp} expression \cite{Karsenty:2009p47916}, PTHrP signaling negatively impacts on its own expression \cite{Kobayashi:2005ge}, which contradicts a key assumption of the model. Moreover, there is no experimental evidence that PTHrP would enhance the removal of IHH; PTHrP rather blocks \textit{Ihh} production by preventing hypertrophic differentiation \cite{Vortkamp:1996wz} and downregulates the action of IHH  \cite{Kobayashi:2005ge}. The reaction kinetics in that model are thus unlikely to reflect the physiological situation. 

A Turing mechanism based on alternative molecular interactions might, however, well underly patterning during long bone development. To that end we have recently shown that the interaction of the Hedgehog protein with its receptor PTCH1 together with the signaling-dependent upregulation of \textit{Ptch1} expression can result in a Schnakenberg-type Turing mechanism \cite{Menshykau:2012kg}.
We showed that this mechanism can explain the observed branching pattern in wildtype and mutant mice. We wondered whether a Turing mechanism based on the Hedgehog-receptor interaction could also explain the emergence of the central-lateral organization in the early bone primordium. Here the situation is somewhat different from the lung in that patterning occurs on a rapidly expanding domain (the growth speed differs greatly between species and bones \cite{Cooper:2013bg}), but the pattern remains stable and no further spots (apart from the secondary growth plate) emerge on the growing domain. This is rather unusual for a Turing pattern, and can be achieved only if feedbacks alter the Turing parameters accordingly without losing the patterning capacity altogether. This is by no means simple, in particular, if considering how small Turing spaces usually are.

In the bone, it is well known how growth and differentiation are controlled by the core regulatory network, i.e. production of the signaling proteins depends on the local density of the particular cell types which changes during long bone development as proliferating chondrocytes concentrate at the ends of the domain and hypertrophic chondrocytes in the center and the entire structure expands.
This allowed us to study a fully coupled growth-signalling model, while in the lung we studied only a  one-way coupled model, i.e. the growth of the domain was prescribed and impacted the signaling, but not vice versa.
The parameter space for which Turing mechanisms yield pattens is typically very small. We therefore wondered whether a model that couples  the IHH-PTCH1-based Schnakenberg-type Turing mechanism with the underlying tissue dynamics could still generate the observed patterns on a growing and differentiating tissue domain, i.e. the emergence of hypertrophic chondrocytes and \textit{Ihh} expression in the center of the domain, the predominance  of proliferating chondrocytes towards the sides of the domain, and the emergence of a differentiation zone towards the center of the domain).

In addition to purely biochemical interactions, mechano-biological cues have been shown to impact chondrocyte proliferation and differentiation, directionality of growth, and tissue deformation of bone development \cite{Henderson2002}.
Furthermore, also the mechanical properties of the surrounding perichondrium impact bone growth by generating tension \cite{Foolen2008}. It was shown though that the direct mechanical impact hypothesis is not sufficient to explain growth regulation \cite{Foolen2009}. Rather, signaling in periochondrial cells is regulated by tension, which in turn indirectly impacts on bone growth \cite{Foolen2011}. Since these mechanisms do not explain the emergence of the central-lateral organization, and because they seem to have an external (not coupled) impact, they are not part of our model. 
\\

In summary, based on the reported biochemical interactions we developed a regulatory model for the core signaling factors IHH, PTCH1, and PTHrP and included two cell types, maturing (proliferating $\&$ resting) chondrocytes as well as pre-hypertrophic chondrocytes. We show that the IHH-PTCH1 module gives rise to a Schnakenberg-type Turing kinetics, and that inclusion of PTHrP is important to achieve robust patterning when coupling patterning and growth.
The model reproduces all relevant spatiotemporal gene expression patterns, as well as  a number of relevant mutant phenotypes. We thus find that a ligand-receptor based Turing mechanism can control the emergence of patterns during long bone development, and that in such a regulatory framework PTHrP is important to confer patterning robustness.

\section*{Model}

Given  the central role of the  IHH/PTHrP feedback loop  in the regulation of endochondral ossification we develop a minimal model that focuses on this IHH-PTHrP feedback loop and the growth of the bone tissue by local proliferation and differentiation (Fig. 1A).
We include three signaling proteins, IHH (symbol for the dimensional concentration $\left[\mathrm{I}\right]$, units $\left[mol/m^{3}\right]$), its receptor PTCH1 (symbol for the dimensional concentration $\left[\mathrm{R}\right]$, units $\left[mol/m^{3}\right]$), and PTHrP (symbol for the dimensional concentration $\left[\mathrm{P}\right]$, units $\left[mol/m^{3}\right]$) and two cell types, proliferating chondrocytes (symbol for the dimensional cell concentration $\left[\mathrm{C}\right]$, units $\left[mol/m^{3}\right]$) and hypertrophic chondrocytes (symbol for the dimensional cell concentration $\left[\mathrm{H}\right]$, units $\left[mol/m^{3}\right]$).
We describe the dynamics of the regulatory signaling proteins and the two cell types, $\mathrm{X}_{i}$, with a set of coupled partial differential equations (PDE) of isotropic advection-reaction-dispersion type,  i.e.
\begin{equation}\label{eq:diffusionequation}
\underbrace{\! \partial_{\overline{t}} [\mathrm{X}_{i}]}_{\text{time derivative}}+ \underbrace{\overline{\nabla} \cdot \left( [\mathrm{X}_{i}] \boldsymbol{\overline{u}}  \right)}_{\text{dilution} \, \& \, \text{advection}} = 
 \underbrace{\overline{D}_{\mathrm{X}_{i}} \overline{\Delta} [\mathrm{X}_{i}]}_{\text{\ diffusion}} +  \underbrace{\mathcal{R}\left([\mathrm{X}_{j}]\right)}_{\text{\ reaction}}
\end{equation}
where $\![\mathrm{X}_{i}]$ denotes the concentration of $\mathrm{X}_{i}$, $\partial_{\overline{t}}$ the time derivative, $\boldsymbol{\overline{u}}$ denotes the external velocity field (units $\left[m/s\right]$),  $D_{\mathrm{X}_{i}}$ represents the diffusion coefficient of component $\mathrm{X}_{i}$ (units $\left[m^{2}/s\right]$), $\overline{\Delta}$ denotes the Laplace operator, and $\mathcal{R}([\mathrm{X}_{j}])$ represents the reaction terms (units $\left[mol/\left(s \cdot m^{3}\right)\right]$) that will be derived.
The term $\overline{\nabla} \cdot \left( [\mathrm{X}_{i}] \boldsymbol{\overline{u}} \right)$ models the effects of dilution and advection as the tissue grows.
The square brackets and overscores denote dimensional variables and operators.
We model tissue as an incompressible fluid and describe tissue growth with a Navier-Stokes equation where the source term depends on the proliferation and differentiation signals. In the following we describe the details of the model.

\subsection*{Tissue Dynamics}

The developing bone structure grows as a result of cell proliferation and differentiation. Proliferating chondrocytes are small in size but expand rapidly while the cell volume of hypertrophic chondrocytes is 4-fold larger than that of resting or proliferating chondrocytes.
The cell density for the two populations is thus different.
Any expansion of these populations in number must thus translate into an expansion of the domain.

The mechanical response of embryonic tissue corresponds to an elastic solid for high frequency perturbations. On a long time scale, on the other hand, the tissue behaves like a viscous fluid, or shows active behavior \cite{Phillips1978}. One possible explanation might be that even when assuming an elastic material, active reorganization of the tissue (proliferation, apoptosis, local cellular rearrangements) leads to a fluidization of the tissue \cite{Ranft2010b}. As a result, a wide range of vertebrate (early) embryonic tissue has previously been shown to be well described by an immiscible viscous fluid on a long time scale \cite{Forgacs1998}. 
We therefore model the mechanical behaviour of the tissue by introducing an incompressible Newtonian fluid with density $\overline{\rho}$ (units $\left[kg/m^{3}\right]$), dynamic viscosity $\overline{\mu}$ (units $\left[kg/\left(s\cdot m\right)\right]$) and local molar source $\mathcal{\overline{S}}$ (units $\left[mol/\left(s\cdot m^{3}\right)\right]$).
The tissue growth model is summarized in Fig. \ref{fig:Fig2}.
This model has been applied to early vertebrate limb development simulations \cite{Dillon:2003p12910} and, in an extended anisotropic formulation, to Drosophila imaginal disc development \cite{Bittig:2008uc}. The Navier-Stokes equation is given as:

\begin{eqnarray}\label{eq:navierstokesequation}
\overline{\rho} \partial_{\overline{t}} \boldsymbol{\overline{u}} &=&
 - \overline{\nabla} \overline{p} + \overline{\mu} \left( \overline{\Delta} \boldsymbol{\overline{u}} + \frac{1}{3}\overline{\nabla} \left( \overline{\nabla} \cdot \boldsymbol{\overline{u}} \right) \right) \\
\overline{\rho} \left( \overline{\nabla} \cdot \boldsymbol{\overline{u}} \right) &=& \overline{\omega} \mathcal{\overline{S}}
\end{eqnarray}

\noindent where
$\overline{p}$ denotes the scalar fluid pressure field (units $\left[kg/\left(m\cdot s^{2}\right)\right]$), and
$\overline{\omega} \mathcal{\overline{S}}= \overline{\omega} (\mathcal{\overline{S}}_{\textit{prol}} + \mathcal{\overline{S}}_{\textit{diff}})$ denotes the local mass production rate, which is composed of contributions from proliferation and increase in cell volume (hypertrophic differentiation).
Assuming dominance of viscous dissipation, we ignored intertial effects (creeping flow).
$\overline{\omega}$ is the cellular mass of proliferating chondrocytes, measured in $\left [\frac{kg}{mol}\right ]$.
For simplicity we did not include Hedghog regulation of chondrocyte proliferation explicitly because its inclusion did not affect the wildtype patterning process in the model. Proliferating chondrocytes thus proliferate at a constant rate $\overline{\varphi}$, i.e.

\begin{equation}\label{eq:sourceprol}
\mathcal{\overline{S}}_{\mathrm{prol}} = \overline{\varphi} [\mathrm{C}]
\end{equation}
Tissue growth by differentiation into hypertrophic chondrocytes is described by:

\begin{equation}
\mathcal{\overline{S}}_{\mathrm{diff}} = \left(\Phi-1\right) \overline{R}_{\mathrm{diff}}  [\mathrm{C}]
\end{equation}
\noindent where $\Phi$ reflects the $\Phi$-fold higher volume of hypertrophic chondrocytes compared to proliferating chondrocytes and $\overline{R}_{\mathrm{diff}} $, as defined in Equation (\ref{eq:Eq_Rdiff}), is the rate of cell differentiation.\\

The interface between the chondrogenic and surrounding tissue is modeled as a passively advected boundary.
Its position, together with the velocity field $\boldsymbol{\overline{u}}$, is coupled to the  morphogen dynamics as described in Equation (\ref{eq:diffusionequation}).\\

The perichondrial bone collar embraces the central anlage, mainly next to the hypertrophic chondrocytes  \cite{Kronenberg:2003p47895, Horton:2009ck}.
The 'Directed Dilation' hypothesis \cite{Wolpert1981,Wolpert1982} states that the perichondrial collar exerts tension and pressure in such a way that radial growth is restricted and the primordium is forced to grow in longitudinal direction.
Therefore, we mimicked the circumferential perichondrium as a tube with rigid walls, which forces the tissue to grow in longitudinal direction.
The length of the tube is chosen so as to permit an about sixfold increase in length as observed in nature \cite{Schipani:2003dh}.
When reaching the ends of the bone collar tube, the tissue forms the characteristic rounded epiphyseal cartilage.
Since the bone collar tube is fixed, it is a non-coupled feature in the model.

\subsection*{The Reaction Kinetics  \label{sec:Kinetics}}

The ligand IHH binds to at least two PTCH1 receptors and forms a multimer \cite{Goetz:2006p49225}.
We therefore use as rate of complex formation $k_\mathrm{on}[ \mathrm{I}] [\mathrm{R}]^2$ ($k_\mathrm{on}$ having units $\left[m^{6}/\left(mol^{2}\cdot s\right)\right]$)
and $k_\mathrm{off} [\mathrm{R}_2 \mathrm{I}]$ ($k_\mathrm{off}$ having units $\left[1/s\right]$)
as rate of dissociation.
\textit{Ihh} is expressed by pre-hypertrophic chondrocytes $\left[\mathrm{H}\right]$ \cite{Bitgood:1995ga, Vortkamp:1996wz}, and we therefore write for the production rate $ \overline{\rho}_{\mathrm{I}} \left[\mathrm{H}\right]$ ($\rho_\mathrm{I}$ measured in $\left[1/s\right]$).
Moreover, we expect that as for other ligands receptor-independent decay of IHH is negligible and we therefore do not include an explicit decay reaction. We then have for the reaction term
\begin{eqnarray} \label{eq:eqI}
\mathcal{R}( [\mathrm{I}])  =  \underbrace{ \overline{\rho}_{\mathrm{I}} [\mathrm{H}]}_{\text{\ production}} \underbrace{-\, \overline{k}_\mathrm{on}[\mathrm{R}]^2[\mathrm{I}] + \overline{k}_\mathrm{off}[\mathrm{R_2I}]}_{\text{\ complex formation}}  
\end{eqnarray}

The expression of the gene \textit{Ptch1} is enhanced by IHH signaling \cite{StJacques:1999vz}, and the rate of \textit{Ptch1} expression must therefore be a function of the concentration of the complex, i.e. $f([\mathrm{R_2I}])$.
We will use a linear approximation as the simplest possible relation for the receptor production rate $f([\mathrm{R_2 I}])$, and write $f([\mathrm{R_2I}]) = \overline{\rho}_\mathrm{R} [\mathrm{C}] + \overline{v} [\mathrm{R_2I}]$, where $\overline{\rho}_{ \mathrm{R}}[\mathrm{C}]$ and $\overline{v}$ are zero and first order rate constants, respectively.
Both $\overline{\rho}_\mathrm{R}$ and $\overline{v}$ are measured in $\left[1/s\right]$.
This linear approximation implies that we assume that the concentration of the IHH-PTCH1 complex  is much lower than the signaling threshold (Hill constant $K$) of a potential cooperative Hill-type regulation of the upregulation of \textit{Ptch1} expression, i.e. $ \frac{[\mathrm{R_2I}]}{[\mathrm{R_2I}]+K} \sim [\mathrm{R_2I}]$.
Unbound PTCH1 is removed by complex formation and restored by its dissociation.
In the absence of contrary experimental evidence we will further assume linear decay of PTCH1 at rate $\overline{\delta}_{\mathrm{R}} [R]$.
Moreover, since \textit{Ptch1} is expressed only on the resting and proliferating, but not on the hypertrophic chondrocytes, PTCH1 will also be lost as a result of cell differentiation at rate $\overline{R}_{\mathrm{diff}}$ (having units $\left[1/s\right]$), which will be defined later in Equation (\ref{eq:Eq_Rdiff}).
The spatio-temporal dynamics of free PTCH1 can then be described by
\begin{eqnarray} \label{eqR}
\mathcal{R}( [\mathrm{R}])  =  \underbrace{ \overline{\rho}_\mathrm{R}[\mathrm{C}] + \overline{\nu} [\mathrm{R_2I]}}_{\text{\ production}} \underbrace{-(\overline{\delta}_{\mathrm{R}} +\overline{R}_{\textit{diff}}) [\mathrm{R}] }_{\text{\ loss}} \underbrace{-2\, \overline{\mathrm{k}}_\mathrm{on}[\mathrm{R}]^2[\mathrm{I}] + \overline{k}_\mathrm{off}[\mathrm{R_2I}]}_{\text{\ complex formation}}  
\end{eqnarray}
The dynamics of the complex $\left[R_2I\right]$ is then described by
\begin{eqnarray} \label{eqPI}
\mathcal{R}( [\mathrm{R^2I}])  = \underbrace{\, \overline{\mathrm{k}}_\mathrm{on}[\mathrm{R}]^2[\mathrm{I}] - \overline{k}_\mathrm{off}[\mathrm{R_2I}]}_{\text{\ complex formation}}  \underbrace{-\overline{\delta}_{\mathrm{R_2I}}[\mathrm{R_2I}]}_{\text{\ degradation}}. 
\end{eqnarray}
where the complex is assumed to be degraded at rate $\overline{\delta}_{\mathrm{R_2I}}[\mathrm{R_2I}]$.

We note that this set of equation corresponds to the classical Schnakenberg Turing model if we make a quasi-steady state approximation for the binding of IHH and PTCH1, while assuming that the diffusion of the membrane-bound IHH-PTCH1 complex is slow compared to its binding and turn-over kinetics. The quasi-steady state concentration of bound receptor, $[\mathrm{R}_2 \mathrm{I}]$,  is then proportional to $[\mathrm{R}]^2\mathrm{[I]}$, i.e.
\begin{eqnarray*} \label{eqPI_qstst}
[\mathrm{R_2I}] _\mathrm{SS}= \frac{\overline{k}_\mathrm{on}}{\overline{k}_\mathrm{off} + \overline{\delta}_{\mathrm{R_2I}}} [\mathrm{R}]^2 [\mathrm{I}] = \overline{\Gamma} [\mathrm{R}]^2 [\mathrm{I}]
\hspace{1cm}\mbox{with}\hspace{1cm}
\overline{\Gamma}= \frac{\overline{k}_\mathrm{on}}{\overline{k}_\mathrm{off} + \overline{\delta}_{\mathrm{R_2I}}}
\end{eqnarray*}
We then have the following equations for the ligand IHH and for the receptor PTCH1:
\begin{eqnarray}\label{eq:nondimIandRreactions}
\mathcal{R}([\mathrm{I}]) &=& \overline{\rho}_{\mathrm{I}} [\mathrm{H}] + ( \overline{k}_\mathrm{off} \overline{\Gamma} -\, \overline{k}_\mathrm{on} ) [\mathrm{R}]^2[\mathrm{I}]   \nonumber \\
\mathcal{R}([\mathrm{R}])  &=& \overline{\rho}_\mathrm{R} [\mathrm{C}]  -(\overline{\delta}_{\mathrm{R}} +\overline{R}_{\textit{diff}}) [\mathrm{R}] + ( (\overline{k}_\mathrm{off}  +\overline{\nu}) \, \overline{\Gamma} -2\, \overline{k}_\mathrm{on}) [\mathrm{R}]^2[\mathrm{I}] 
\end{eqnarray}

The gene \textit{Pthrp} is expressed by non-hypertrophic resting or proliferating chondrocytes $\left[\mathrm{C}\right]$ in response to signaling of the IHH-PTCH1 complex $\left[\mathrm{R}^2\mathrm{I}\right]$ \cite{Karsenty:2009p47916}.
For simplicity, we use a Heaviside function, $\mathcal{H}\left(\cdot\right)$,  to implement the threshold response of \textit{Pthrp} expression to IHH signaling.
$\left[P\right]$ is produced when the complex concentration $[\mathrm{R}]^2[\mathrm{I}]$ drops below the threshold $\overline{\theta}_{\mathrm{P}}$.
When then have:
\begin{eqnarray} \label{eq:eqPTHrP}
\mathcal{R}( [\mathrm{P}])  &=& \overline{\rho}_P  \mathcal{H}\left( [\mathrm{R}]^2[\mathrm{I}]- \overline{\theta}_{\mathrm{P}} \right) [\mathrm{C}] - \overline{\delta}_{\mathrm{P}} [\mathrm{P}]
\hspace{1cm}
\mbox{with}
\hspace{1cm}
\mathcal{H}\left(x\right) =
\begin{cases} 0 &\mbox{if } x < 0 \\
1 & \mbox{if } x \geq 1 \end{cases} 
\end{eqnarray}
%

\subsection*{Cell Dynamics}
Both proliferating chondrocytes $\left[\mathrm{C}\right]$ and pre-hypertrophic chondrocytes $\left[\mathrm{H}\right]$ are represented as cell number densities and obey Equation (\ref{eq:diffusionequation}). The dilution term for a component $\mathrm{X}$ in Equation (\ref{eq:diffusionequation}) is given by $[\mathrm{X}] \left( \overline{\nabla} \cdot \overline{\boldsymbol{u}} \right)$. The divergence of the velocity field is composed of the contributions of proliferation and differentiation:
\begin{equation}
\overline{\nabla} \cdot \boldsymbol{\overline{u}}  = 
\frac{\overline{\omega}}{\overline{\rho}} \left(  \mathcal{\overline{S}}_{\mathrm{prol}} + \mathcal{\overline{S}}_{\mathrm{diff}} \right) =
\frac{\overline{\omega}}{\overline{\rho}}  \left( \overline{\varphi} + \left(\Phi-1\right)  \overline{R}_{\mathrm{diff}} \right) [\mathrm{C}]
\end{equation}
where $\overline{\varphi}$, measured in $\left[1/s\right]$, represents the constant proliferation rate (c.f. Equation (\ref{eq:sourceprol})).
Proliferating chondrocytes differentiate at rate  $\overline{R}_{\mathrm{diff}}$ into hypertrophic chondrocytes. Upon proliferation, the cell density of the proliferative chondrocytes $[\mathrm{C}]$ increases with the rate $\overline{\omega}/\overline{\rho} \overline{\mathcal{S}}_{prol}$. We thus obtain:
\begin{eqnarray}
\mathcal{R}\left([\mathrm{C}]\right) &=& - \overline{R}_{\mathrm{diff}} [\mathrm{C}] +
\frac{\overline{\omega}}{\overline{\rho}} \overline{\mathcal{S}}_{prol} [\mathrm{C}] \\
\mathcal{R}\left([\mathrm{H}]\right) &=& \Phi \overline{R}_{\mathrm{diff}} [\mathrm{C}]
\end{eqnarray}

The diffusion constants of the cells are small and equal, i.e. $\overline{D}_{\mathrm{C}}=\overline{D}_{\mathrm{H}}=\overline{D}_{\mathrm{cell}}$. With Equation (\ref{eq:sourceprol}), we get the following partial differential equations:
\begin{eqnarray} \label{cells}
\partial_{\overline{t}} [\mathrm{C}] + 
\overline{\nabla} \cdot \left( [\mathrm{C}] \overline{\boldsymbol{u}} \right) & = &
\overline{D}_\mathrm{cell} \overline{\Delta} [\mathrm{C}] - \overline{R}_{\mathrm{diff}} [\mathrm{C}] +
\frac{\overline{\omega}}{\overline{\rho}} \overline{\varphi} [\mathrm{C}]^{2}
\nonumber\\
\partial_{\overline{t}} [\mathrm{H}] +
\overline{\nabla} \cdot \left( [\mathrm{H}] \overline{\boldsymbol{u}} \right) &=& 
\overline{D}_\mathrm{cell} \overline{\Delta} [\mathrm{H}]+ \Phi \overline{R}_{\mathrm{diff}} [\mathrm{C}] 
\end{eqnarray}

We will study two different models of cell differentiation. In the simpler model there is a direct coupling between IHH signaling and cell differentiation.
In the second model cell differentiation is controlled by the protein PTHrP which in turn is regulated by IHH signaling
(i.e. by $[\mathrm{R}]^2[\mathrm{I}]$).
For the directly coupled model (index \textit{d}), the local differentiation sets in when the local IHH signaling $\mathrm{R}^2\mathrm{I}$ deceeds the threshold $\overline{\theta}_{d}$.
The indirectly coupled model (index $i$), on the other hand, leads to differentiation once PTHrP, $\mathrm{P}$, deceeds the threshold $\overline{\theta}_{i}$:

\begin{subequations}\label{eq:Eq_Rdiff}
\begin{align}
    \overline{R}_{\mathrm{diff}} ^{d} 	&=\overline{\delta} \, \mathcal{H}\left(\overline{\theta}_{d}-[\mathrm{R}]^2[\mathrm{I}] \right) 	\label{eq:rdiff_direct} \\
    \overline{R}_{\mathrm{diff}} ^{i} 	&=\overline{\delta} \, \mathcal{H}\left(\overline{\theta}_{i}-[\mathrm{P}]\right) 			\label{eq:rdiff_indirect}
\end{align}
\end{subequations}
Here $\bar{\delta}$ denotes the differentiation rate; $\mathcal{H}\left(\cdot\right)$ denotes the Heaviside function.\\

\subsection*{Boundary Conditions}

Indian Hedgehog $\left[\mathrm{I}\right]$ can diffuse freely into the  tissue that surrounds the nascent bone domain, e.g. the limb. The surrounding tissue is of finite size $L_{\mathrm{domain}}$, and zero-flux boundary conditions are applied for $\left[\mathrm{I}\right]$ at the edge of the surrounding domain.
PTCH1 $\left[\mathrm{R}\right]$, PTHrP $\left[\mathrm{P}\right]$,  and the cells $\left[\mathrm{C}\right]$ and $\left[\mathrm{H}\right]$, are restricted to the chondrogenic tissue of length $L_{\mathrm{bone}}$, i.e. zero-flux boundary conditions $\boldsymbol{n} \cdot \overline{\nabla} [\mathrm{X}]=0$ are applied to $\left[\mathrm{R}\right]$, $\left[\mathrm{P}\right]$, $\left[\mathrm{C}\right]$ and $\left[\mathrm{H}\right]$.
In case of the receptor PTCH1 this reflects its membrane-bound state. In case of PTHrP this is a good approximation given its low diffusion coefficient.

\subsection*{Parameter Values}
The measurement of the parameter values (i.e. diffusion coefficients, production and degradation rates)  \emph{in vivo} is complicated and has only been carried out in few model systems, but not in the developing bone \cite{Kicheva2007, Yu2009, Ries2009}. However, our conclusions do not depend on the exact values of parameters, but mainly depend on their relative values as can be seen by non-dimensionalizing the model. To non-dimensionalize the model we need to choose characteristic length and time scales, as well as characteristic concentrations. As characteristic time scale we use $T$.
Accordingly we have to rescale our dimensional time $\overline{t}$ into its non-dimensional counterpart $t$, i.e. $\overline{t} = T t$.
As characteristic length scale we use $L$.
Accordingly we have to rescale our dimensional spatial coordinates $\overline{x}_{i}$ into their non-dimensional counterparts $x_{i}$, i.e. $\overline{x}_{i} = L x_{i}$, and similar for the velocity field $ \boldsymbol{\overline{u}} =\boldsymbol{u} \frac{L}{T} $.
We non-dimensionalize the Ihh, Ptch1, and PTHrP concentrations with respect to their characteristic concentrations, i.e. $\mathrm{I} = [\mathrm{I}]/\mathrm{I}_0$, $\mathrm{R} =  [\mathrm{R}]/ \mathrm{R}_0$, and $\mathrm{P} =  [\mathrm{P}]/ \mathrm{P}_0$.
Similarly for the cells we use the characteristic densities, i.e. $\mathrm{C} =  [\mathrm{C}]/ \mathrm{C}_0$, $\mathrm{H} =  [\mathrm{H}]/ \mathrm{H}_0$.

The non-dimensional parameters are obtained by non-dimensionalizing with the respective scales, i.e. $\delta = T \overline{\delta}$.
The equations can be further simplified by choosing appropriate scales and by combining parameter combinations into new composite parameters.
Thus we write for the characteristic time scale $T = \frac{L^2}{\overline{D}_\mathrm{R} }$.
This choice of time scale leads to the classic Schnakenberg equations with the diffusion coefficient of the slowly diffusing factor being equal to 1.
The characteristic length scale is set to $L = \sqrt{\gamma \overline{D}_\mathrm{R} / \overline{\delta}_\mathrm{R}}$ with the non-dimensional scaling factor $\gamma = T 
\overline{\delta}_{\mathrm{R}} =  \frac{L^2 \overline{\delta}_{\mathrm{R}}}{\overline{D}_\mathrm{R}}$.
The fluid pressure $\overline{p}$ is non-dimensionalized as $\overline{p} = \frac{\overline{\mu}}{T} p$.
The dimensionless Reynolds-number $Re=\frac{\overline{\rho} L^{2}}{\overline{\mu} T}$ is a measure for the relative importance of inertia over viscous effects.
For the characteristic protein concentrations we use $\mathrm{R}_0 = \sqrt{\frac{\overline{\delta}_{\mathrm{R}}}{ \overline{k}_\mathrm{on} -\overline{k}_\mathrm{off} \overline{\Gamma} }}$,
$\mathrm{I}_0 = \frac{1}{\mathrm{R}_0} \frac{\overline{\delta}_{\mathrm{R}}}{2\, \overline{k}_{\mathrm{on}} - (\overline{k}_{\mathrm{off}}  +\overline{\nu}) \, \overline{\Gamma}}$, and $\mathrm{P}_0 = \frac{\overline{\rho}_{\mathrm{P}}}{\overline{\delta}_{\mathrm{R}}}$.
Finally for the characteristic cell densities we set $\mathrm{C}_0 = \mathrm{H}_0 = \frac{\overline{\rho}}{\overline{\omega}}$. 

We obtain as composite parameters the non-dimensional diffusion constants $D_{\mathrm{I}} = \frac{\overline{D}_\mathrm{I}}{\overline{D}_\mathrm{R}}$,
$D_{\mathrm{P}} = \frac{\overline{D}_\mathrm{P}}{\overline{D}_\mathrm{R}}$, $D_{\mathrm{cell}} = \frac{\overline{D}_\mathrm{cell}}{\overline{D}_\mathrm{R}}$,
the differentiation rate $\delta =  \frac{\overline{\delta}}{\overline{\delta}_{\mathrm{R}}} $,
the proliferation rate $\varphi  =  \frac{\overline{\varphi}}{\overline{\delta}_{\mathrm{R}}} $,
and the production rates  $\rho_\mathrm{I} = \frac{\overline{\rho}_{\mathrm{I}}}{\overline{\delta}_{\mathrm{R}}}  \frac{\mathrm{H}_0}{\mathrm{I}_0} $,
$\rho_R = \frac{\overline{\rho}_{\mathrm{R}}}{\bar{\delta}_R R_0} $,
and the thresholds
$\theta_{\mathrm{P}} = \frac{\overline{\theta}_{\mathrm{P}}}{\mathrm{R}_{0}^{2} \mathrm{I}_{0} }$,
$\theta_{d} = \frac{\overline{\theta}_{d}}{\mathrm{R}_{0}^{2} \mathrm{I}_{0}}$,
$\theta_{i} = \frac{\overline{\theta}_{i}}{\mathrm{P}_{0}}$.
The non-dimensional model reads:
\begin{eqnarray}\label{eq:cde_system_nondim}
\partial_{t} \mathrm{I} +  \nabla \cdot (\mathrm{I} \boldsymbol{u} )  &=& D_{\mathrm{I}} \Delta \mathrm{I}  +
\gamma \left( \rho_{\mathrm{I}} \mathrm{H} -
\mathrm{R}^2 \mathrm{I}  \right) \nonumber \\
\partial_{t} \mathrm{R} +  \nabla \cdot (\mathrm{R} \boldsymbol{u} ) &=& \Delta \mathrm{R}  +
\gamma \left(
 \rho_{\mathrm{R}} \mathrm{C} -
\left(1+ R_{\mathrm{diff}}\right)  \mathrm{R} +
\mathrm{R}^2\mathrm{I}
\right)\nonumber \\
\partial_{t} \mathrm{P} + \nabla \cdot (\mathrm{P} \boldsymbol{u} )  &=& D_{\mathrm{P}}  \Delta \mathrm{P} +
\gamma \left( \mathcal{H}\left( \mathrm{R}^2\mathrm{I}- \theta_{\mathrm{P}} \right) \mathrm{C}
- \delta_{\mathrm{P}} \mathrm{P}
\right)
\nonumber \\
\partial_{t} \mathrm{C} +  \nabla \cdot \left( \mathrm{C} \boldsymbol{u} \right) &=& D_{\mathrm{cell}}  \Delta \mathrm{C} -
\gamma \left( R_{\mathrm{diff}} \mathrm{C} +
\gamma \varphi \mathrm{C}^{2}
\right)
\nonumber \\
\partial_{t} \mathrm{H} + \nabla \cdot \left( \mathrm{H} \boldsymbol{u} \right) &=& D_{\mathrm{cell}} \Delta \mathrm{H}  +
\gamma \Phi R_{\mathrm{diff}} \mathrm{C}
\end{eqnarray}
\begin{subequations}\label{eq:r_diff_dimless}
\begin{align}
R_{\mathrm{diff}}^{d} &= \delta \mathcal{H}\left(\theta_{d} - \mathrm{R}^{2}\mathrm{I} \right) \label{eq:r_diff_dimless_direct} \\
R_{\mathrm{diff}}^{i} &= \delta \mathcal{H}\left(\theta_{i} - \mathrm{P}\right) \label{eq:r_diff_dimless_indirect}
\end{align}
\end{subequations}
\begin{subequations}\label{eq:navierstokesequation_nondim}
\begin{align}
 Re  \partial_{t} \boldsymbol{u}  &=
 - \nabla p + \Delta \boldsymbol{u} + \frac{1}{3}\nabla \left( \nabla \cdot \boldsymbol{u} \right) \\
\nabla \cdot \boldsymbol{u} &=   \gamma (\left(\Phi-1\right) R_{\mathrm{diff}}  + \varphi) \mathrm{C} \label{eq:navierstokesequations_continuity_nondim}
\end{align}
\end{subequations}

The dimensionless model contains seven parameters less than the dimensional one and the patterning mechanism no longer depends on absolute diffusion and decay constants, but only on the relative diffusion coefficients and the relative decay rates.
Similarly, the absolute protein concentrations do not matter but only the relative concentrations (as a result from the relative expression  and decay rates) relative to the threshold concentrations.
The non-dimensional parameter values are given in Table \ref{tab:parameters} and lie within the Turing space.
Here we used a 10-fold higher diffusion rate for the ligand $\mathrm{I}$ than for its receptor $\mathrm{R}$ which is in good agreement with experimental observations   \cite{Kumar2010, Hebert2005,Kicheva:2008p5770, Swaminathan:1997p20237, Yu:2009p22070}.
Moreover, the proliferation and differentiation rate are set equal in the model and are likely to be of similar order of magnitude in reality.
The production rates are difficult to compare, given the different normalizations, and depend on the signaling thresholds $\theta$. \\

\subsubsection*{Initial Values}
The initial values of all components are summarized in Table \ref{tab:parameters}.
We use zero concentration for $\mathrm{I}$ though also non-zero concentrations could be used without affecting the pattern.
The initial $\mathrm{P}$ concentration is chosen such that differentiation is blocked initially.
$\mathrm{P}$ is degraded and diluted due to the expansion of the primordium, such that differentiation begins at later times when the $\mathrm{P}$ concentration has dropped below the critical concentration at which it prevents a cell from differentiating.
Additionally to the initial $\mathrm{P}$ concentration, the influx of maternal $\mathrm{P}$ may support the initial inhibition of differentiation.
PTCH1 is internalized constitutively and can be visualized at the cell surface only by blocking or significantly delaying its internalization \cite{Incardona:2000p14489}.
Although PTCH1 is not detected at significant levels on the cell surface at steady state, even these low levels rapidly remove IHH from the cell surface \cite{Incardona:2000p14489}, presumably through rapid cycling at rate.
We therefore use a non-dimensional concentration of one as initial condition. \\

Initially there are mainly proliferating chondrocytes in the digit primordia and we thus use as nondimensional cell densities $\mathrm{C}_{0}=0.9$ and $\mathrm{H}_{0}= 0.1$ for proliferating and hypertrophic chondrocytes, respectively. \\

The chondrogenic tissue of initial non-dimensional length $L_{\mathrm{bone}}(t=0)=1$ and width $0.4$ was placed in the center of a square of non-dimensional size $L_{\mathrm{domain}}=20$, representing the surrounding tissue.\\

\subsubsection*{Comparison of Model Parameter Values to Measured Values}
The parameter values can in part be compared to experimental values by converting these to their dimensional counterparts based on the established length and developmental time scales. Concentration ranges, on the other hand, have not been determined experimentally. \\

The growth speed differs between bones. Murine embryonic tibia bones measure about 1 mm at embryonic day $14$ and grow by about $250\,\mu m$ between embryonic day 14 and 15, and faster thereafter \cite{Blakley:1979uh}. Chicken femurs measure about $5\,mm$ at embryonic day 10 and grow by about $1\,mm$ until day 11 and take about 2.5 days to double in length\cite{Kanczler:2012ez}.
Thus in both embryos the growth accelerates, but this acceleration appears to be mainly driven by ossification in the center, and would thus not apply to our model which focuses on an earlier process.
Since we did not find any data for this earlier growth process we will use the earliest known growth rates.
In the simulation, the tissue grows about two-fold in the interval $t=\left[0,20\right]$ which would correspond to about 4 days in case of the murine measurements and 5 days in case of the chicken.
$t = 1$ thus corresponds to $\overline{t} = 5\,\left[days\right] = 18000\,\left[s\right]$ and the characteristic time scale is thus $T \approx 18000\,\left[s\right]$.
We then have $\overline{\delta}_{\mathrm{R}} = \gamma/T \approx 5.6 \times 10^{-3}\,\left[s^{-1}\right]$ which is some 100-fold higher than the typical ligand-independent receptor turn-over rates that have been measured for other receptor systems (in the order of $5\times10^{-5}\,\left[s^{-1}\right]$) \cite{Eden2011,Lloyd:1983vc, Valley:2008hp,Woods:1989ui,Ginsburg:1998tl,Singh:2001tx} and almost 10-fold higher than the ligand-induced Dpp receptor turn-over rate ($5\times 10^{-4}\,\left[s^{-1}\right]$) \cite{Mizutani:2005p943}.
While this may appear high,  Ptch is well known to be internalized constitutively at a rather high rate such that blockage of internalization greatly increases its membrane concentration \cite{Incardona:2000p14489}.
Besides the embryonic growth rate may well be lower initially which would correspond to a lower value for  $\overline{\delta}_{\mathrm{R}}$. \\

The characteristic length is $L = \sqrt{T \overline{D}_\mathrm{R}}$.
The diffusion coefficient for membrane receptors is in the range $D=0.001\dots0.1\,\left[\mu m^2 s^{-1}\right]$ \cite{Kumar2010, Hebert2005}.
Using the upper value of $\overline{D}_\mathrm{R} = 0.1~\left[\mu m^2 s^{-1}\right]$ we obtain as characteristic length scale  $L = 42~\left[\mu m\right]$.
The initial non-dimensional length of the simulated bone structure is $L_{\mathrm{bone}}(t=0) = 1$ which then corresponds to $42~\left[\mu m\right]$ and may well reflect the initial size of the mesenchymal condensations from which the bone forms by endochondral ossification.
Again if the embryonic growth rate was lower initially this would correspond to larger size of the initial mesenchymal condensations.
For the other diffusion coefficients we would have  $\overline{D}_{\mathrm{I}} = 10\times\overline{D}_{\mathrm{R}} = 1 \left[\mu m^2 s^{-1}\right]$ which is well within the physiological range for soluble proteins such as Dpp and Wg in the wing disc, if somewhat on the high end \cite{Kicheva:2008p5770, Swaminathan:1997p20237, Yu:2009p22070}.
Finally, for PTHrP we have  $\overline{D}_{\mathrm{P}} = 0.01 \times \overline{D}_{\mathrm{R}} =  0.001~\left[\mu m^2 s^{-1}\right]$ which implies that PTHrP has to diffuse rather poorly for the patterning mechanism to work. This restriction could be removed by introducing the PTH/PTHrP receptor as a further variable. For the sake of parsimony we refrain from doing this.  \\

The dynamic viscosity of tissue is known to be approximately $\overline{\mu} \approx 10^{4}~\left[Pa\cdot s\right]$ \cite{Forgacs:1998fy}, some $10^7$-fold higher than for water, and the mass density of the tissue is $\overline{\rho} \approx 1000 \left[kg/m^{3} \right]$.
The Reynolds number computes as $Re = \frac{\overline{\rho} L^{2}}{\overline{\mu} T} \approx 10^{-14}$ and is virtually zero.
As expected the inertial effects are neglectably small (Stokes regime) and the terms on the left hand side of Equation (\ref{eq:navierstokesequation_nondim}) can therefore be omitted.

\subsection*{Simulation}
The PDEs were solved with finite element methods as implemented in COMSOL Multiphysics 4.3a. COMSOL Multiphysics is a well-established software package and several studies confirm that COMSOL provides accurate solutions to reaction-diffusion equations both on constant\cite{cutress2010} and growing two-dimensional domains\cite{Carin2006, Thummler2007, Weddemann2008, Germann:bT_kMV7D, Germann:uu}. The mesh and the time step were refined until further refinement no longer resulted in noticeable improvements as judged by eye.

For Equations (\ref{eq:cde_system_nondim}), the \textit{Coefficient Form PDE} module with \textit{Zero Flux}
boundary condtions have been used both for the moving boundary and the boundary of the outer domain.
The dilution $\mathrm{X}_{i} \left( \nabla\cdot \boldsymbol{u}\right)$ term is taken as a right hand side
contribution $f=-\mathrm{X}_{i} \mathcal{S}$.

To solve Equations (\ref{eq:navierstokesequation_nondim}), namely Equation (\ref{eq:navierstokesequations_continuity_nondim}),  a local mass source has been
added to the weak expressions of the \textit{Creeping Flow} module. The intertia terms in Equation (\ref{eq:navierstokesequation})
have been omitted, as well as $\mu/3 \nabla \left(\nabla \cdot \boldsymbol{u} \right)$, since these terms are
negligible for our sets of parameters.

The \textit{Moving Mesh} ALE module was applied to the mesh. The moving boundary's mesh velocity was set to the 
local fluid velocity, and \textit{Prescribed Mesh Displacement} was used for the bone collar.

The Heaviside function $\mathcal{H}\left(\cdot\right)$ is smoothed sigmoidally in the interval $[-0.05,0.05]$.\\

\section*{Results}

\begin{figure}[h!]
\begin{centering}
\includegraphics[width=0.8\columnwidth]{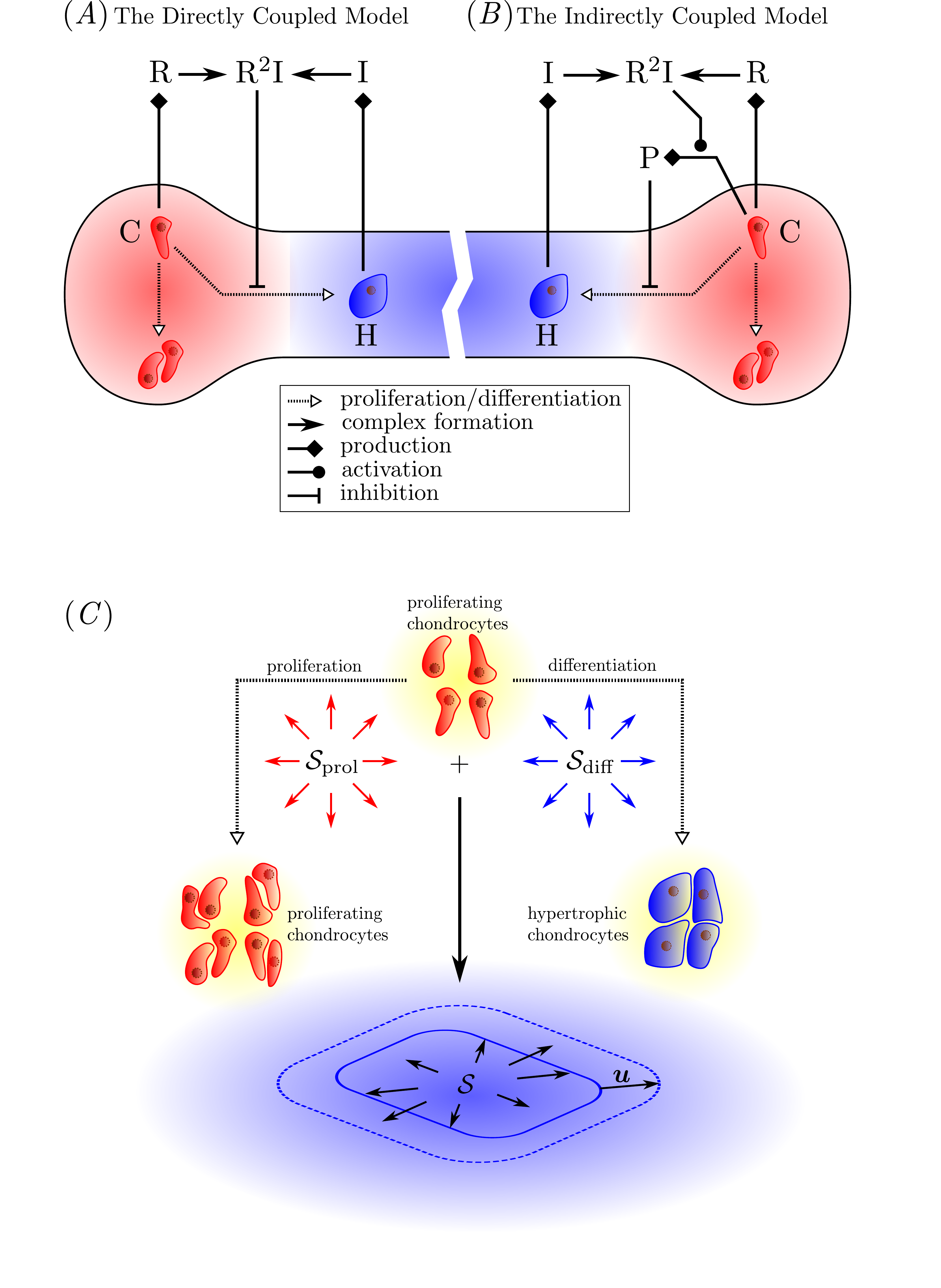}
\par\end{centering}
\caption{}
\label{fig:Fig2}
\end{figure}

In a first step, we solve the signaling network on a constantly growing domain, where only proliferation is considered (referred to as the \textit{prescribed growth model}). We demonstrate that the Turing pattern has the potential to control the central-lateral organization on a growing domain. In a second model, differentiation is directly regulated by the level of IHH signaling (referred to as the \textit{directly coupled model}, see Fig. \ref{fig:Fig2} (\textit{A})). We find that the direct impact of the Turing system on differentiation leads to  volatility in the patterning. In a final model, differentiation is indirectly controlled via the level of a paracrine factor that inhibits cell differentiation, and which in turn is controlled by the level of IHH signaling (referred to as the \textit{indirectly coupled model}, see Fig. \ref{fig:Fig2} (\textit{B})). This factor can be interpreted as PTHrP. By introducing this additional integrator we retrieve stability and reproduce all experimentally observed patterns under wildtype and mutant conditions. 

\subsection*{Patterning on a bone tissue that expands according to a prescribed overall growth rate}

We were first interested whether the signaling mechanism could at all result in the experimentally observed patterning.
To avoid the complication of the feedbacks between patterning and tissue growth we first solved the signaling system on a constantly growing domain, i.e. without IHH signaling affecting growth and differentiation (and thus the local mass source $ \mathcal{S}$).
There is thus no spatially varying feedback on the production as would arise from the (here ignored) cell differentiation;  the pattern thus evolves for constant (Turing) parameter values. Growth is prescribed by a function. In a first study we analyse linear growth and the local proliferation rate must therefore decreases as the total area $A\left(t\right)$ of the tissue increases, i.e.

\begin{equation}
 \mathcal{S}_{\mathrm{prol}}\left(t\right) = \gamma \varphi  \frac{A\left(t=0\right)}{A\left(t\right)} \mathrm{C}.
\end{equation}

\noindent The overall growth rate of the domain is then constant, corresponding to an approximately constant growth speed of the long bone. This is equivalent to a growth rate proportional to a passively advected and diluted compound. To capture the experimental growth pattern \cite{Schipani:2003dh} we adjusted the length of the bone collar tube such that the  domain would grow about fourfold longitudinally before growing radially to form the epiphysis. Growth in width is minor and can be controlled by the distance of the bone collar to the chondrogenic tissue. \\

\begin{figure}
\begin{centering}
\includegraphics[width=1\columnwidth]{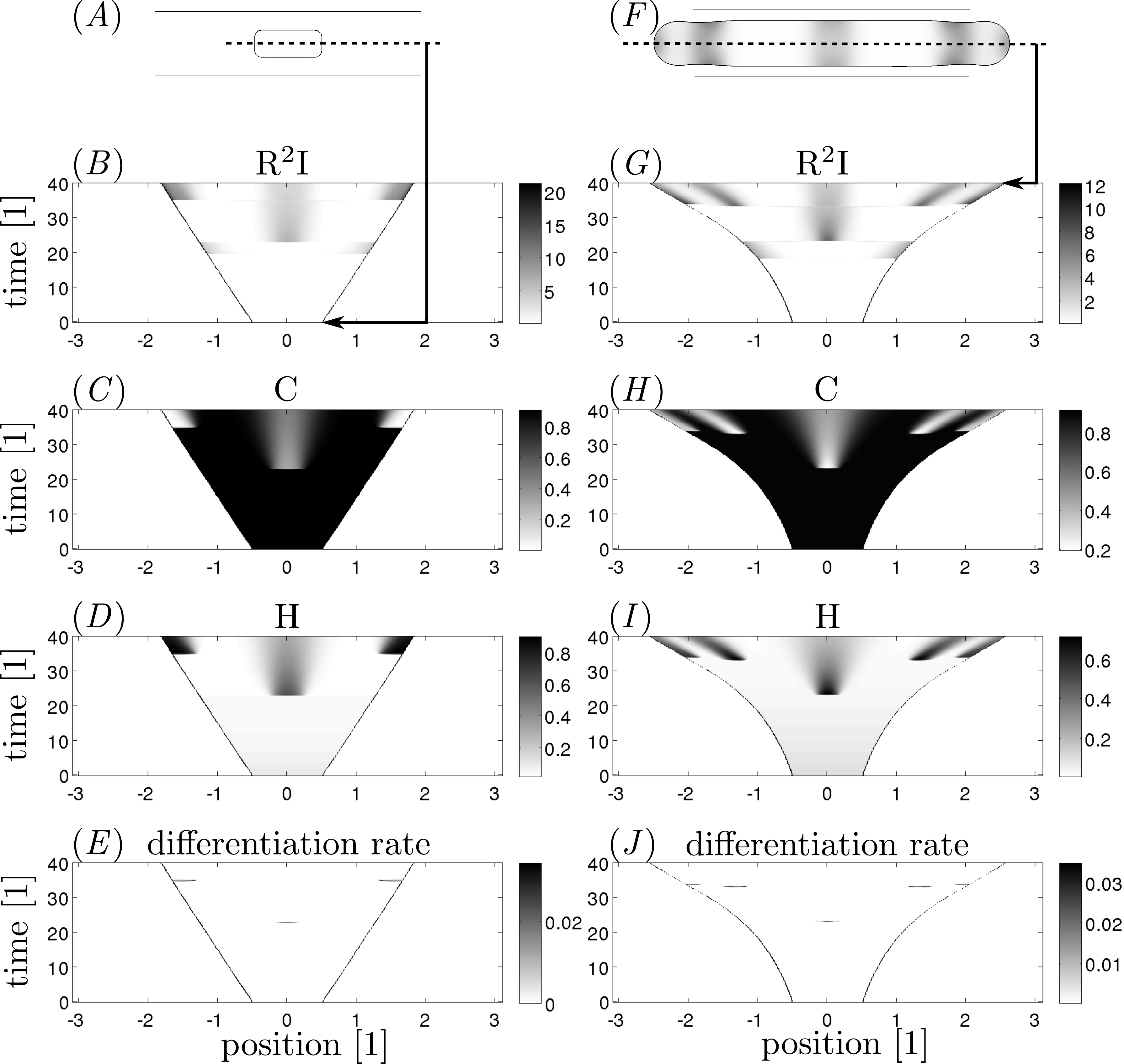}
\par\end{centering}
\caption{}
\label{fig:Fig3}
\end{figure}

Figure \ref{fig:Fig3} (\textit{A}) shows the initial domain at  $t=0$.
The other panels Figure \ref{fig:Fig3} (\textit{B-E}) show how the patterns of various components evolve over time (until $t=40$) along the midline that is indicated in Figure \ref{fig:Fig3} (\textit{A}).
We focus on the pattern in the middle of the domain because most pattern changes occur along the length of the domain, while the pattern is relatively stable and uniform along the width of the domain. 

Thus Figure \ref{fig:Fig3} (\textit{B}) shows the concentration of the signaling complex $\mathrm{R}^2\mathrm{I}$ on the midline.
The Turing patterning mechanism, based on the interaction between $\mathrm{I}$ and $\mathrm{R}$, results in spatial patterns as the domain reaches a critical size at $t \approx 17.7$, and $\mathrm{R}^2\mathrm{I}$ concentrates at the edges of the domain (Figure \ref{fig:Fig3} (\textit{B})).
This is consistent with the experimental observation that IHH signaling induces \textit{Pthrp} expression in that part of the domain. IHH signaling ($\mathrm{R}^2\mathrm{I}$) is present throughout the domain, though at a much lower concentration. This would explain the observed impact of IHH signaling on the proliferation of chondrocytes.

The Turing pattern emerges at time $t \approx 19.7$ and stays stable until $t \approx 23.1$, when a pattern inversion occurs (Figure \ref{fig:Fig3} (\textit{B})) and high IHH signaling also emerges in the centre of the domain.
However, at that stage many additional processes have already started (most importantly the invasion of osteoblasts and ossification) that affect the patterning and that are not part of our model which focuses on the initial patterning events.
At later times $t \approx 35.2$ the high IHH signaling reappears at the ends.
Whenever IHH signaling is absent or low for long times, the initially nonzero PTHrP concentration decreases and finally drops below the threshold where proliferating chondrocytes start to differentiate.
This occurs at time $t\approx 22.9-23.1$ (central differentiation) and $t \approx 34.7-36.3$ (lateral differentiation; c.f. Figure \ref{fig:Fig3} (\textit{E})).
As a result proliferation (and resting) chondrocytes (pC) are reduced in the center of the domain (Figure \ref{fig:Fig3} (\textit{C})) and hypertrophic chondrocytes emerge (Figure \ref{fig:Fig3} (\textit{D})).
The transition between the two populations are sharper in the embryo, but overall the signaling mechanism appears to produce realistic patterns with the linear prescribed domain growth.
\\

We also analysed pattern formation on an exponentially growing domain, i.e.
\begin{equation}
 \mathcal{S}_{\mathrm{prol}}\left(t\right) = \gamma \varphi  \mathrm{C}.
\end{equation}
An example of the geometry at a later time point ($t=40$) is shown in Figure \ref{fig:Fig3} (\textit{F}).
In the initial phase we observe similar patterning dynamics as with linear growth because the growth rate is small and in the same range.
At later stages the growth speed increases significantly, leading to different dynamics of the Turing system.
It is known from the analysis of Turing models on exponentially growing domains that the number of spots correlates with the size of the domain \cite{Crampin:1999p46562}.
This property is also found on our dynamically growing domain (Figure \ref{fig:Fig3}(\textit{F-G})).
Interestingly, the emerging patterns are not evenly spaced.

\subsection*{Direct coupling of Turing Patterns and Growth result in pattern instability}

We next solved a model in which growth and patterning are directly coupled via signaling of the IHH-receptor complex. The differentiation rate  $R_{\mathrm{diff}}^{d}$  is now directly controlled by $\mathrm{R}^2\mathrm{I}$ (Equation (\ref{eq:r_diff_dimless_direct})).

\begin{figure}
\begin{centering}
\includegraphics[width=1\columnwidth]{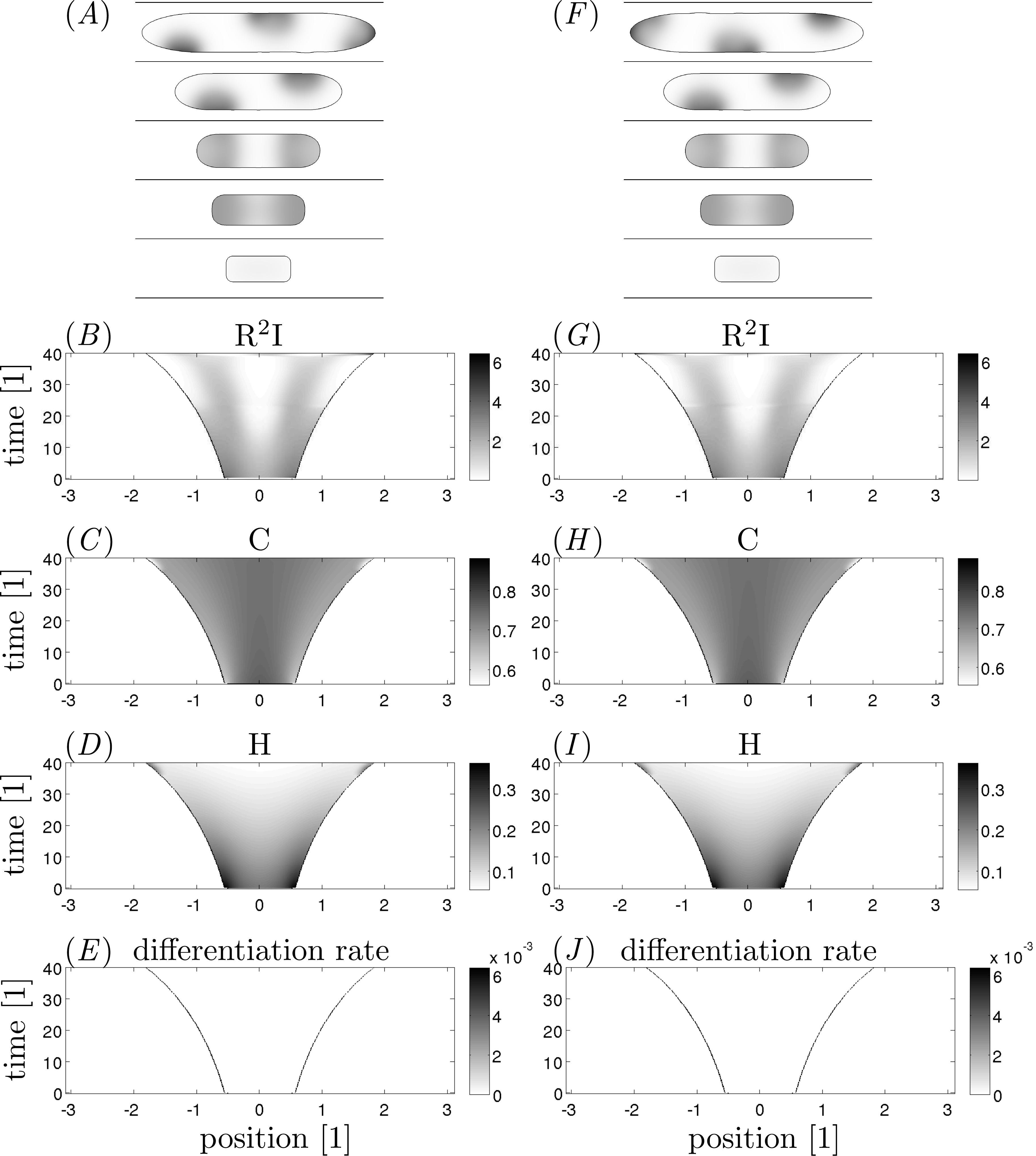}
\par\end{centering}
\caption{}
\label{fig:Fig4}
\end{figure}

Since the signal $\mathrm{R}^2\mathrm{I}$ is zero at the beginning, differentiation occurs immediately and leads to a spatially homogeneous loss of the proliferating chondrocytes and an accumulation of hypertrophic chondrocytes, respectively (Figure \ref{fig:Fig4} (\textit{C-E})).
The hypertrophic chondrocytes start to express \textit{Ihh}, which, in turn, activates its signaling pathway.
The Turing pattern emerges at $t\approx 0.4$, but cannot rescue a spatial organization any more due to the high levels of \textit{Ihh} expression.
Due to the uniformly inhibited differentiation, the hypertrophic chondrocyte population is only diluted as a result of proliferation.
At time $t\approx 22.8$, the radial symmetry is lost and the Turing system does not lead to the correct cellular organization any more.

The high sensitivity on initial values of this model is another feature arguing against its biological relevance.
Figure \ref{fig:Fig4} (\textit{F-J}) show the same simulation as shown in Figure \ref{fig:Fig4} (\textit{A-E}), but with slightly disturbed initial cell concentrations (uniform distribution with mean 1 and range 0.05).
Although the initial system dynamics is stable, the perturbations of the initial conditions lead to different solutions at later stages as shown in Figure \ref{fig:Fig4} (\textit{A}) and (\textit{F}).
This behaviour is a strong indication for implausibility of the model.

As expected for a Turing pattern, the length of the bone primordium determines the number of spots (or modes) (Figure \ref{fig:Fig4}), i.e. the distance between each two spots cannot exceed a certain length, otherwise a new spot emerges in between.
The patterning dynamics are very fast because diffusion and reactive turnover dominate over advective transport for our choice of parameters.
This pattern transition behaviour conforms to the well studied Turing system behaviour on domains with prescribed growth \cite{Crampin:1999p46562}.

\subsection*{Stable Patterning with an indirect coupling of patterning and growth}

Instead of the IHH-PTCH1 complex, $\mathrm{R}^2\mathrm{I}$, controlling differentiation directly, we now introduce PTHrP production by proliferative chondrocytes, $\mathrm{P}$, as an intermediator. The IHH-PTCH1 complex, $\mathrm{R}^2\mathrm{I}$, induces expression of this intermediator $\mathrm{P}$, which can diffuse and prevents differentiation into hypertrophic chondrocytes as described by $R_{\mathrm{diff}} ^i$ in Equation (\ref{eq:r_diff_dimless_indirect}).  \\

\begin{figure}
\begin{centering}
\includegraphics[width=1\columnwidth]{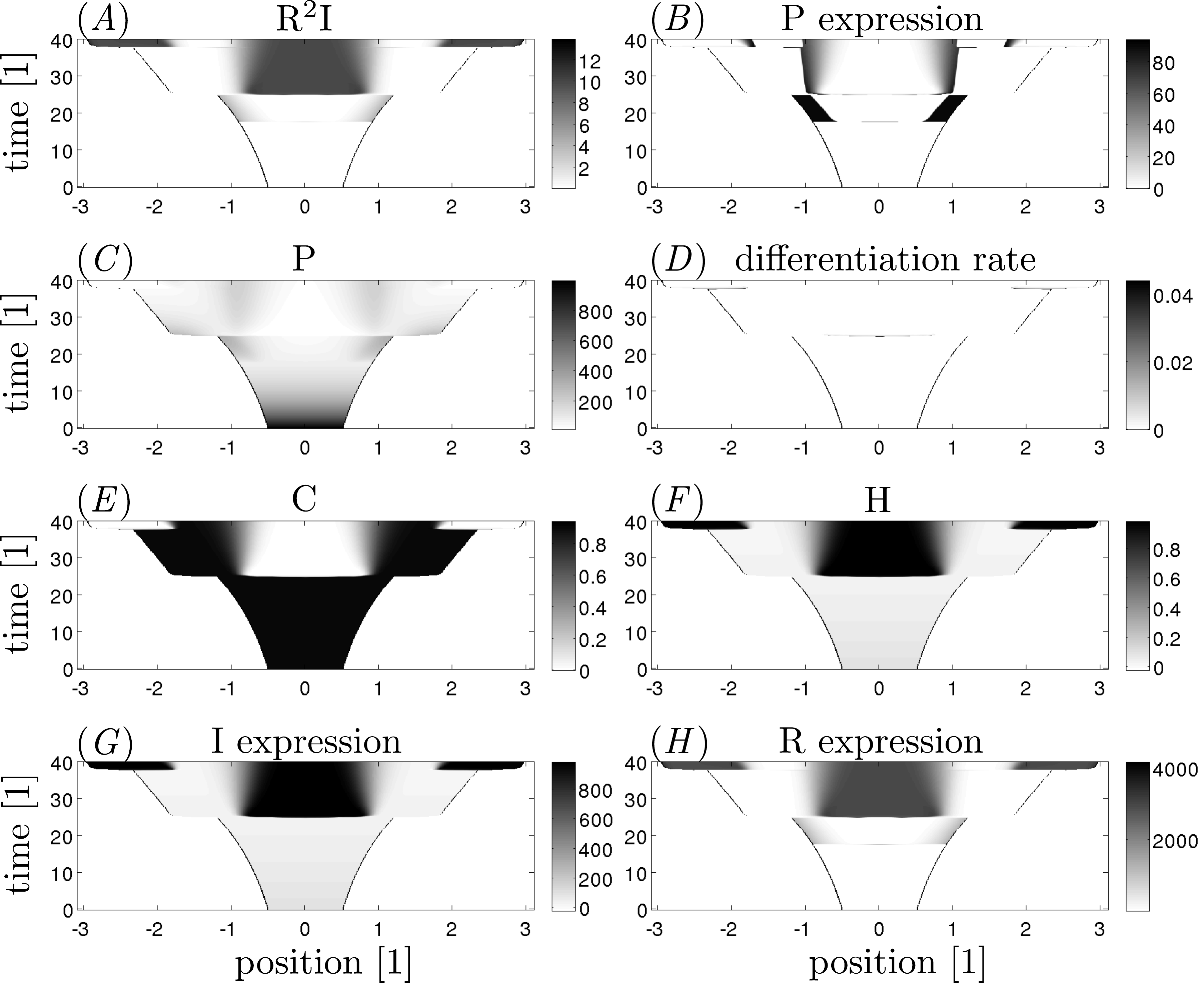}
\par\end{centering}
\caption{}
\label{fig:Fig5}
\end{figure}

The IHH-PTCH1 complex, $\mathrm{R}^2\mathrm{I}$, emerges at time $t\approx 17.7$ and stays stable until $t\approx 24.9$ (Figure \ref{fig:Fig5} (\textit{A})).
The complex accumulates at the ends of the domain, which triggers expression of \textit{Pthrp} (Figure \ref{fig:Fig5} (\textit{B})).
As the pattern changes at time $t\approx 24.9$, \textit{Pthrp} expression moves slightly towards the center of the domain as indeed observed in the embryo.
While PTHrP accumulates at the ends, where it prevents differentiation into hypertrophic chondrocytes, it is further degraded in the centre of the domain, and finally drops below the threshold of differentiation inhibition (Figure \ref{fig:Fig5} (\textit{C})).
Differentiation starts at time $t\approx 24.9$ (Figure \ref{fig:Fig5} (\textit{D})) and leads to fast transition of proliferating into hypertrophic chondrocytes (Figure \ref{fig:Fig5} E-F) and therefore to high growth rates.
The high \textit{Ihh} expression by (pre-)hypertrophic chondrocytes in the centre (cf. Figure \ref{fig:Fig5} (\textit{G})) subsequently leads to high IHH signaling (\ref{fig:Fig5} (\textit{A})), since the residual proliferating chondrocytes are still able to express \textit{Ptch1} at high levels.

\begin{figure}
\begin{centering}
\includegraphics[width=1\columnwidth]{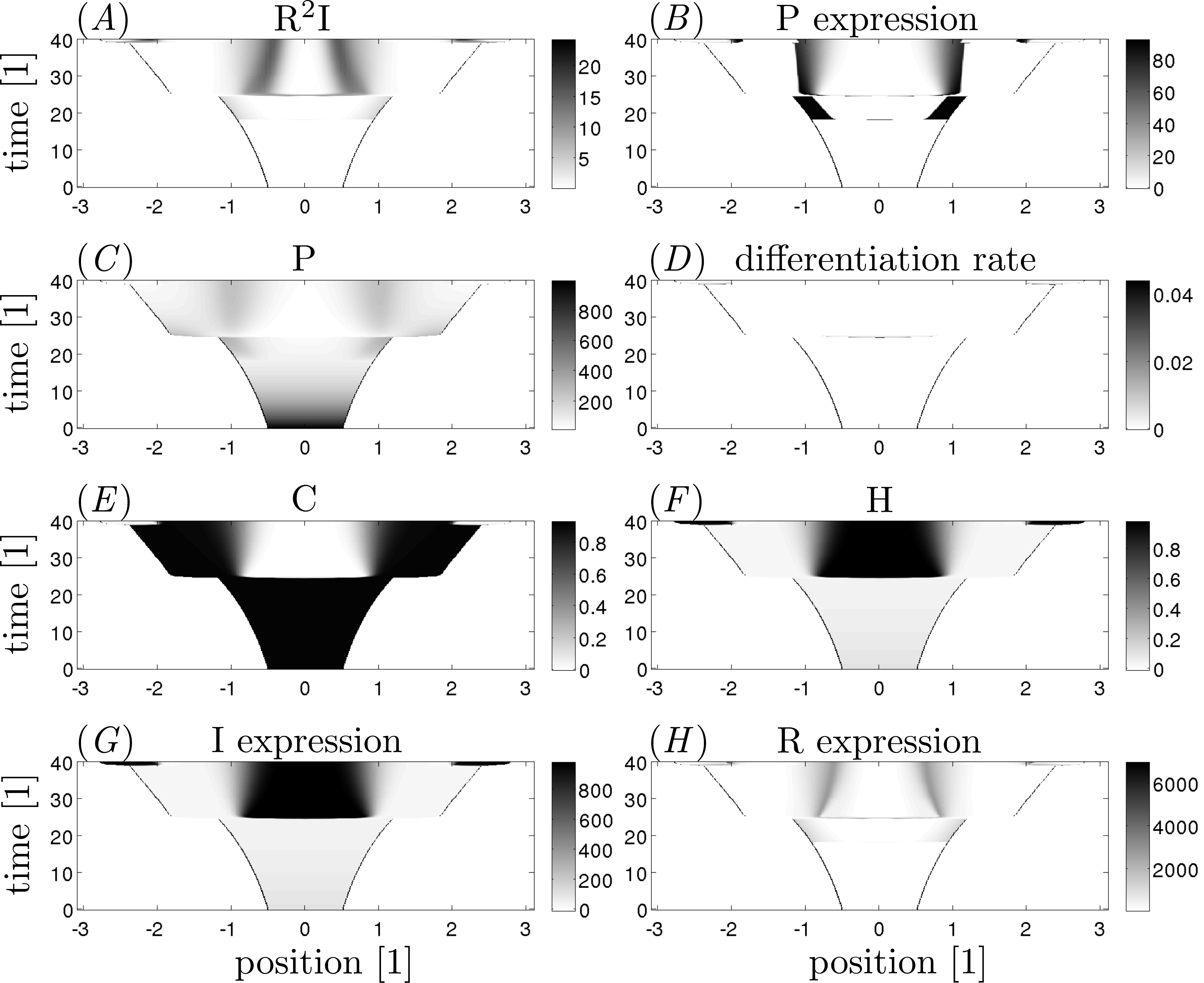}
\par\end{centering}
\caption{}
\label{fig:Fig6}
\end{figure}

Experiments also reveal the IHH-dependent expression of \textit{Ptch1} adjacent to the \textit{Ihh}-expressing domain in the center \cite{StJacques:1999vz}.
This pattern has so far been counterintuitive given that IHH signaling induces \textit{Pthrp} expression at the very ends of the domain.
In the model the rate of \textit{Ptch1} expression is given by $\overline{\rho}_\mathrm{R} [\mathrm{C}] + \overline{\nu} [\mathrm{R}^2 \mathrm{I}]$ where the first term captures the constitutive expression and the second term the effect of the signaling-dependent positive feedback.
If the signal-induced \textit{Ptch1} production is much smaller than the generation of PTCH1 by ligand-unbinding and recycling (i.e. $\overline{\nu} \ll \overline{k}_{\mathrm{off}}$) then we indeed observe \textit{Ptch1} expression adjacent to the \textit{Ihh} expressing domain (Figure \ref{fig:Fig5} (\textit{H})).
There is indeed good experimental evidence for strong PTCH1 recycling which would correspond to the large $\overline{k}_{\mathrm{off}}$  \cite{Incardona:2000p14489}. \\
In order to avoid very high IHH signaling at places where the proliferative chondrocyte density is very low, we introduced a threshold on IHH signaling-induced \textit{Ptch1} expression by proliferating chondrocytes (Figure \ref{fig:Fig6}). 
The expression term reads $\gamma \left( \rho_{\mathrm{R}} \mathrm{C}  + \mathrm{R}^{2}\mathrm{I} \mathcal{H}\left(\mathrm{C}-0.05\right)  \right)$,
where the threshold value $0.05$ is chosen such that the \textit{Ptch1} expression is largely removed from the center of the domain.
As a result, the IHH signaling is restricted to the transition zone between proliferative and hypertrophic chondrocytes (Figure \ref{fig:Fig6} (\textit{A})),
and so is the \textit{Ptch1} expression (Figure \ref{fig:Fig6} (\textit{H})).
The \textit{Pthrp} expression (Figure \ref{fig:Fig6} B), differentiation pattern (Figure \ref{fig:Fig6} (\textit{D})), cell organization 
(Figure \ref{fig:Fig6} (\textit{E-F})) and \textit{Ihh} epxression (Figure \ref{fig:Fig6} (\textit{G})) are only marginally affected.
\\

With the domain growing further, the IHH signaling decreases at the ends and differentiation - analogously to the central differentiation - begins at time $t\approx37.7$.
Again, the high differentiation rate leads to a high growth speed, resulting in bulged structures mimicking the epiphyses.
The accumulation of hypertrophic chondrocytes can be interpreted as the secondary ossification centers.
The hypertrophic chondrocytes at the tips produce \textit{Ihh}, leading to a significant upregulation of IHH signaling (Figure \ref{fig:Fig6} (\textit{A})) and \textit{Ptch1} expression (Figure \ref{fig:Fig6} (\textit{H})).
We note, however, that this model may not be adequate to describe the processes in the epiphysis.
The here described patterning corresponds to a first phase during endochondral ossification.
In a second phase apoptosis of hypertrophic chondrocytes and invasion of osteoblasts from the perchondrium drive bone growth \cite{Schipani:2003dh}.
Proliferation and differentiation of chondrocytes continues in the two separated domains. This part of the process is not included in this model as we were interested in the mechanism by how symmetry is broken and by which the differentiation pattern is first generated. \\

\subsubsection*{Mutants}
\begin{figure}
\begin{centering}
\includegraphics[width=1\columnwidth]{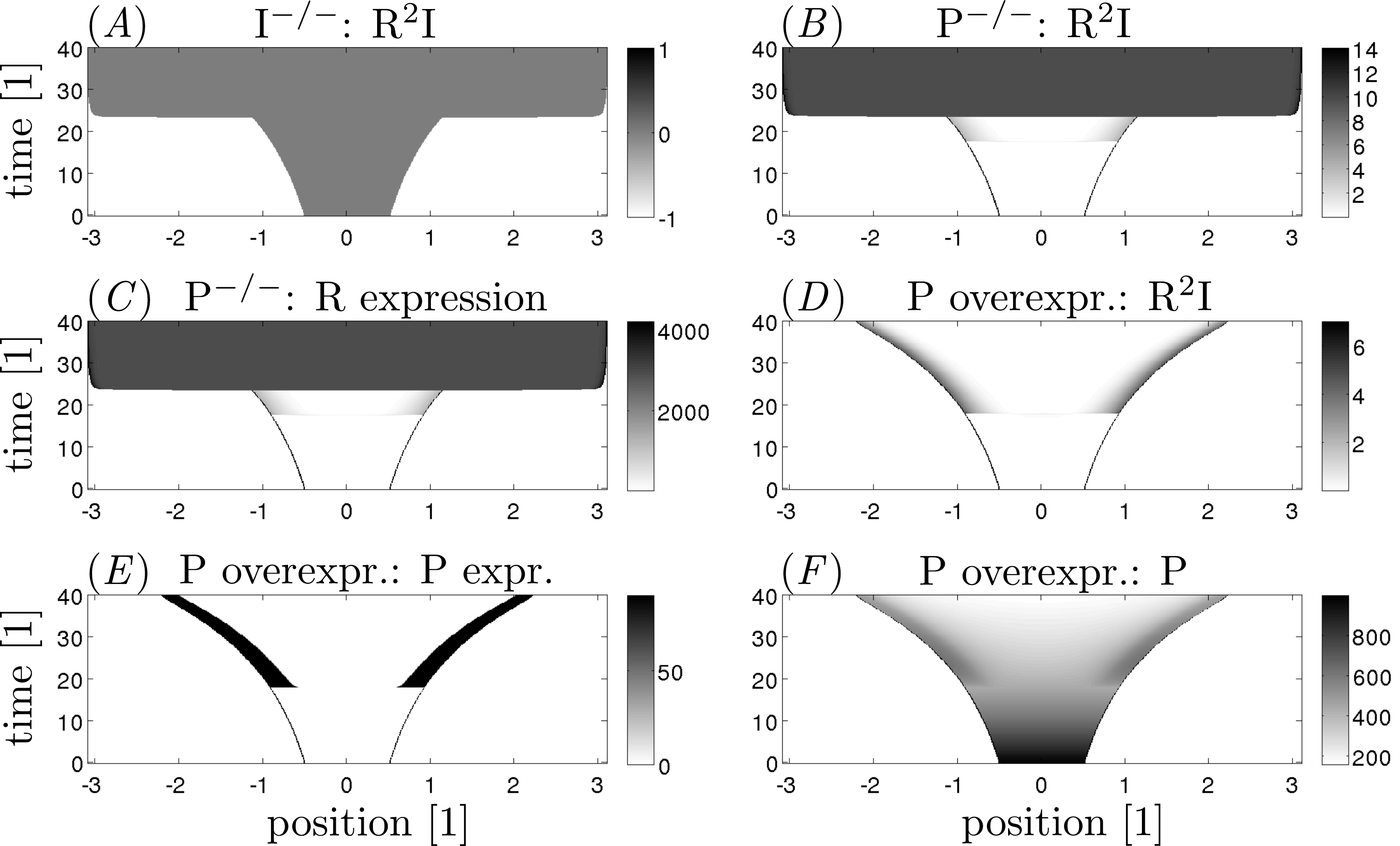}
\par\end{centering}
\caption{}
\label{fig:Fig7}
\end{figure}

An important test for the suitability of a mathematical model is its consistency with a wide range of independent experimental
observations.
Knock-out mice exist for all proteins included in the model.
In the \textit{Ihh} null mouse (\textit{Ihh}$^{-/-}$) only a reduced number of hypertrophic chondrocytes accumulate in the center of the domain, but premature chondrocytes no longer vanish from the center and no functional bone forms \cite{StJacques:1999vz}.
The gene expression pattern in the \textit{Ihh} null mouse (\textit{Ihh}$^{-/-}$) are greatly disturbed: \textit{Ptch} and \textit{Pthrp} expression are absent, and the PTHrP receptor is misexpressed \cite{StJacques:1999vz}.
In our model, IHH signaling is indeed completely absent, and the complex is homogeneously zero (Figure \ref{fig:Fig7} (\textit{A})).
Consequently, the distribution of premature chondrocytes remains spatially homogenous.
However, the degradation and dilution of PTHrP leads to uniform differentiation at time $t\approx 23.3$ and thus to instantaneous growth.
This is not observed in the embryo, presumably because of the differentiation blocking effects, that are not included in this simple model.
%
%

Bones in mice lacking Parathyroid hormone (PTH) and PTHrP are much smaller than in the wildtype, and the zones of proliferating chondrocytes and bone formation are shrunk \cite{Miao:2002dt}. Hypertrophic chondrocytes, however, still emerge in the center of the domain, perhaps because of the effects of maternal PTHrP given the much more severe phenotype in \textit{Pth/Pthrp receptor}$^{(-/-)}$ mutant mice \cite{Lanske:1996vc}.
Most \textit{Pth/Pthrp receptor}$^{(-/-)}$ mutant mice die and those that survive exhibit accelerated differentiation of chondrocytes in bone, and their bones, grown in explant culture, are resistant to the effects of PTHrP and Sonic hedgehog \cite{Lanske:1996vc}.
In our model, the spatial distribution of IHH signaling emerges correctly in the beginning (cf. Figure \ref{fig:Fig7} (\textit{B-C})), but spatially organized cell organization is absent.
Similar to the \textit{Ihh}$^{-/-}$ simulation, the uniform differentiation due to degradation and dilution of the initially available PTHrP concentration of maternal origin, leads to high growth rates.

Overexpression of \textit{Pthrp} leads to blocking of hypertrophic differentiation and to delayed endochondral processes \cite{Weir1996}.
Indeed, differentiation is completely absent in our simulation, although the IHH signaling and the downstream \textit{Pthrp} expression pattern form correctly (cf. Figure \ref{fig:Fig7} (\textit{D-E})).
The PTHrP concentration levels are consistently high (Figure \ref{fig:Fig7} (\textit{F})).

\subsubsection*{Sensitivity $\&$ Robustness}
\begin{figure}
\begin{centering}
\includegraphics[width=1\columnwidth]{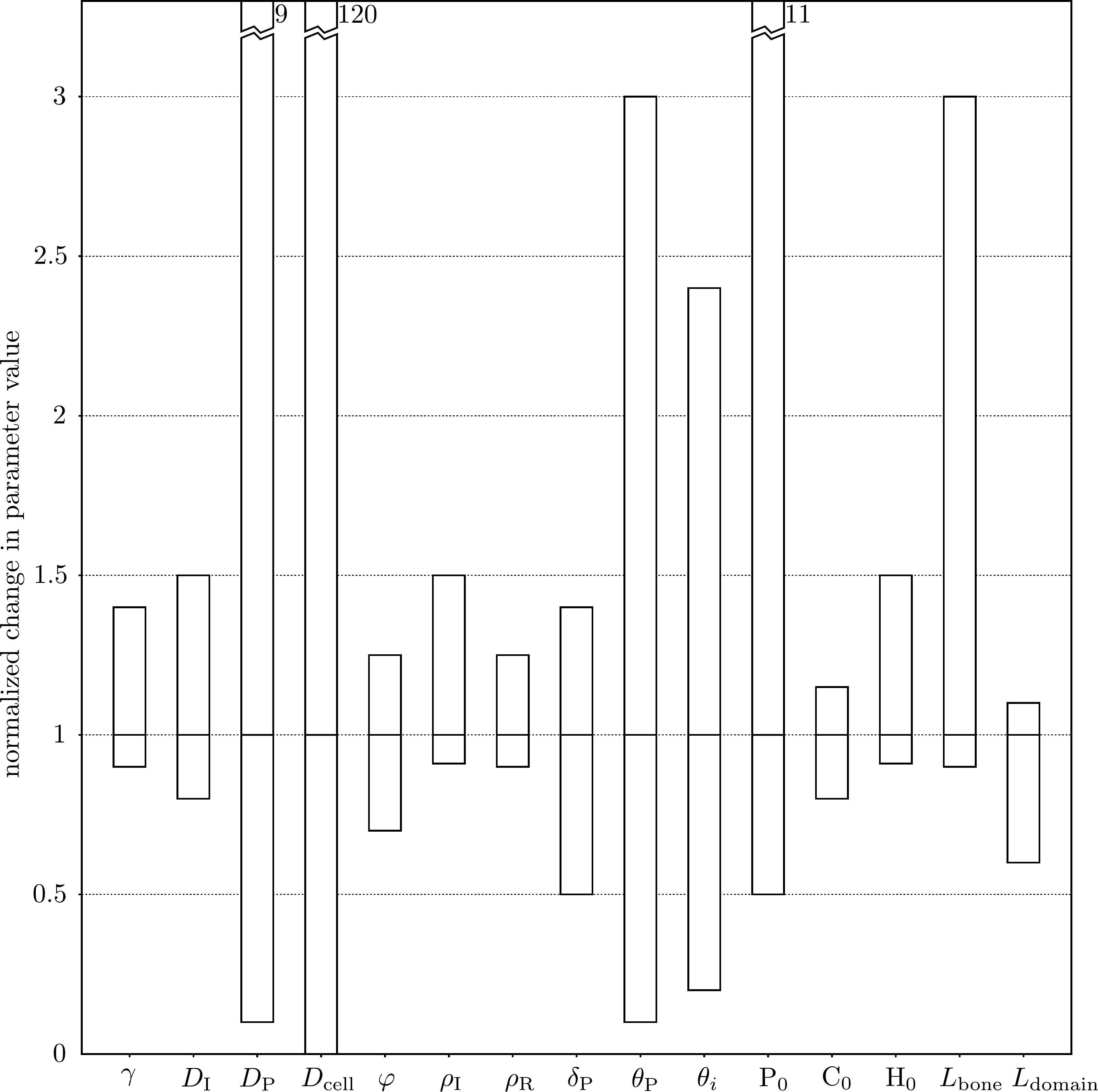}
\par\end{centering}
\caption{}
\label{fig:Fig8}
\end{figure}

When probing the robustness of the model by varying the setpoint parameters we notice large differences in the impacts of the parameter values (Figure \ref{fig:Fig8}).
Those parameters that define the Turing space have the largest impact on the patterning process.
Most importantly, these are the initial proliferative and hypertrophic cell densities $\mathrm{C}_{0}$ and $\mathrm{H}_{0}$, and the expression rates of $\mathrm{I}$, $\rho_{\mathrm{I}}$, and $\mathrm{R}$, $\rho_{\mathrm{R}}$, as these together define the initial total \textit{Ihh} and \textit{Ptch} expression rates $\rho_{\mathrm{I}} \mathrm{H}_{0}$ and $\rho_{\mathrm{R}} \mathrm{C}_{0}$, repectively (cf. Equation (\ref{eq:cde_system_nondim})).
Accordingly, the system is highly sensitive to these parameters.
Furthermore, the Schnakenberg system depends on the ratio of the diffusion coefficients $D_{\mathrm{I}}$ and $D_{\mathrm{R}}$, as well as on the scaling factor $\gamma$.
While $D_{\mathrm{R}}=1$ is kept constant, the diffusion coefficient of IHH, $D_{\mathrm{I}}$, can only be modestly varied.
Similar considerations apply to the scaling factor $\gamma$.
The diffusion coefficient of PTHrP and the cells can be varied massively without affecting the qualitative emergence of the central-lateral organization.
The threshold parameters $\theta_{\mathrm{P}}$ (controlling the \textit{Pthrp} expression) and $\theta_{i}$ (controlling the differentiation) can be varied by almost one order of magnitude.
This is a consequence of the high ratio of minimal and maximal values in the complex pattern, which is one of the characteristic features of Turing patterns.
The initial PTHrP concentration $\mathrm{P}_{0}$ affects the time point when differentiation starts.
Is it too low, differentiation may set in when the complex pattern is not yet formed.
Higher values, on the other hand, lead to delayed differentiation, but still lead to qualitatively correct phenotypes.
Besides dilution, the initial decrease of PTHrP is governed by the protein degradation $\delta_{\mathrm{P}}$, whose value can be changed moderately.
The system is relatively sensitive to changes in the proliferation rate $\varphi$.
This is due to the resulting exponential longitudinal growth in the beginning.
With insufficient proliferation, the primordium is not able to reach the critical length where the pattern appears.
Too high proliferation, on the other hand, prevents the pattern to form.
Here also the initial length of the bone, $L_{\mathrm{bone}}\left(t=0\right)$, is important, and in particular shorter initial lengths are not tolerated very well.
For a shortened initial geometry, the constant proliferation of the proliferating chondrocytes leads to an initial lengthening until the critical aspect ratio is reached, at which the Turing pattern emerges. However, if the initial geometry is too short, proliferation cannot reach the critical length any more within the period under consideration.
The size of the outer domain, $L_{\mathrm{domain}}$, mimicking the limb bud ectoderm, can be changed moderately.
%

We notice that patterning is much less sensitive to the choice of parameters for PTHrP than for those affecting the Turing module directly.
We conclude that the Turing system is relatively sensitive to changes in parameters, whereas, in particular, the differentiation process is highly robust to perturbations.

\section*{Discussion}

We developed a model for long bone development based on the known regulatory interactions. In line with our previous work \cite{Menshykau:2012kg} we find that the IHH-PTCH1 interaction can result in a Schnakenberg-type Turing system. We further show that this patterning module results in the experimentally observed cell distributions as well as gene expression and signaling patterns if the equations are solved on a bone-shaped domain that expands at the embryonic growth speed.
Thus hypertrophic chondrocytes accumulated in the center of the domain and expressed \textit{Ihh} while proliferating chondrocytes accumulated to the sides and expressed \textit{Ptch}.
In the embryo patterning and growth are linked and we therefore also coupled our Turing model with a tissue mechanics model in a next step.
Here the IHH-PTCH complex inhibited differentiation to hypertrophic chondrocytes and thus promoted expansion of proliferating chondrocytes.
Since hypertrophic chondrocytes expressed \textit{Ihh}, the tissue dynamics affects the expression of the Turing components.
Given the small size of the parameter space for which Turing patterns can be observed we were concerned that the coupling would destroy the pattern.
Indeed when we coupled the patterning module directly to the tissue mechanics module the pattern became highly unstable, and we were no longer able to find parameter sets that would give realistic patterns. 

In a next step we introduced a further component, PTHrP. IHH binding to PTCH1 induces the expression of \textit{Pthrp}, and PTHrP in turn inhibits the differentiation to hypertrophic chondrocytes. With this intermediate step added we could now obtain realistic patterns also when the Turing system was coupled to the tissue mechanics module. We reproduced all experimentally observed cell distributions as well as gene expression and signaling patterns while the bone domain was growing at the experimentally observed growth speed.
In addition, \textit{Pthrp} expression was restricted to the edges of the domain as observed in the embryo \cite{Hilton:2005cu}. 

The model was non-dimensionalized before analysis to reduce the number of parameters. As far as quantitative measurements are available (growth speed, domain size, diffusion coefficients) we find that the model parameter values are consistent with measured values when transformed back to their dimensional counterparts. Moreover, the model reproduces the observed gene expression patterns as well as key mutant phenotypes.
Further quantitative comparisons and data-based parameter estimation procedures as are common for ordinary differential equation-based models \cite{Geier:2012gl,Raol:2004p928}, however, are not yet possible, given the available data.

While the good match of model and experimental observations supports a central role of IHH signaling in the early patterning during long bone formation we note that the \textit{Ihh}$^{-/-}$ \textit{Gli3}$^{-/-}$ double knock-out exhibits normal early patterning (with some later defects), and \textit{Pthrp} expression remains restricted to the ends of the domain \cite{Hilton:2005cu, Koziel:2005jr}. This observation is at odds with our current model, as GLI3 is a key downstream transcription factor of IHH signaling. In the absence of IHH, the receptor PTCH1 remains unbound and in its unbound form PTCH1 enhances the expression of the \textit{Gli3} gene \cite{Butterfield:2009p39700} and the proteolytic processing of the GLI3 protein into the GLI3R repressor form, which prevents expression of downstream target genes such as \textit{PthrP} \cite{Hilton:2005cu, Koziel:2005jr}. As a result, \textit{Ihh} null mice cannot express \textit{Pthrp}, and the mesenchymal condensations differentiate uiformly into hypertrophic chondrocytes \cite{StJacques:1999vz, Hilton:2005cu, Koziel:2005jr}. In the absence of \textit{Ihh} and \textit{Gli3}, we would expect spatially uniform expression of \textit{PthrP}. The spatial restriction of \textit{PthrP} expression in \textit{Ihh}$^{-/-}$ \textit{Gli3}$^{-/-}$ double knock-out mice therefore points to an alternative patterning process that restricts \textit{Pthrp} expression to the sides and thereby restricts the emergence of hypertrophic chondrocytes to the center.

One possible mechanism by which \textit{Pthrp} expression may be restricted to the ends of the domain are factors that are secreted by the joints. Various ligands from the TGF-$\beta$ family are all present in the joints and SMAD3-dependent signaling and signals of BMPR-IA (ALK3) have previously been shown to stimulate \textit{Pthrp} expression \cite{Pateder:2001dd,Pateder:2000kb,Zhang:2003gh,Zou:1997wm}. Conditional removal of \textit{Smad4} with \textit{Hoxa13-cre} prevents cartilage formation and ossification \cite{Benazet:2012ewa}.
Inclusion of such a IHH-independent cue of \textit{Pthrp} expression on the distal boundaries would trivially result in the observed distal restriction of the \textit{Pthrp} expression patterns. Future modelling efforts will be focused on understanding the self-organized emergence of BMP signalling that marks the joints. Moreover, we plan to carefully evaluate the contribution of the BMP-dependent expression of \textit{Pthrp} to the wildtype patterning process based on published mutant phenotypes.

In summary, the model reproduces experimental data very well and provides an explanation for the formation of the counterintuitive pattern that requires long-distance signaling of the morphogen IHH.
The ligand-receptor based Turing mechanism can yield such pattern using physiological parameter values. PTHrP had to be introduced in the model as a mediator between the Turing system and the tissue growth module to ensure the stability of the pattern on the expanding domain. Further experimental studies are required to define the redundant or parallel mechanisms that ensure the correct expression  of \textit{Pthrp} in \textit{Ihh}$^{-/-}$ \textit{Gli3}$^{-/-}$ double knock-out mice and to quantify their relative contributions in the wildtype.
BMP-receptor signaling provides a good candidate mechanism.
Interestingly, we have previously shown that also BMP-receptor interactions can give rise to Schnakenberg-type Turing patterns \cite{Badugu:2012ho}.\\

Turing mechanisms have been proposed for many patterning phenomena. Yet, the Turing components have  been difficult to define. The example of long bone patterning adds another example, where a  ligand-receptor based Turing mechanisms can describe both the wildtype and mutant conditions very well, even when embedded in a feedback that controls the expansion of a rapidly growing tissue. This supports the notion that Turing mechanisms in nature may be more generally implemented based on a receptor-ligand interaction. For receptor-ligand interactions to give rise to Turing patterns, ligand and receptors need to interact cooperatively (non-linearly), the receptor-ligand interaction must increases the receptor concentration on the membrane, and ligands need to diffuse faster than receptors. These three conditions are met by many receptor-ligand systems, such that this mechanism could be widely used in nature.

\section*{Acknowledgment}

The authors acknowledge funding from the SNF Sinergia grant "Developmental engineering of endochondral ossification from mesenchymal stem cells" and are grateful for discussions with Vaclav Klika and with members of the consortium and of  the CoBi and Vortkamp groups.

\section*{Author Contributions}
DI and ST developed the model. ST analysed the model. DI and ST wrote the paper.

\section*{References}

\bibliographystyle{plain}
\bibliography{Library_Papers}

\clearpage
\newpage

\section*{Figure Legends}

\paragraph{Figure 1: Signaling in Long Bone Development.} 
Hypertrophic chondrocytes secret the protein IHH (arrow (A)1), and proliferating chondrocytes express the IHH receptor PTCH1 (gene \textit{Ptch1}) (A2), as well as the diffusible, extracellular protein PTHrP (gene \textit{Pthrp}) (A3). IHH signaling, which results from the binding  of IHH to its receptor PTCH1, enhances PTHrP synthesis (A4). PTHrP production is also stimulated by BMP signaling (A5). PTHrP inhibits the differentiation of proliferating chondrocytes into hypertrophic chondrocytes (A6). The proliferation rate of the proliferating chondrocytes is enhanced by IHH signaling (A7).

\paragraph{Figure 2: Modeled Long Bone Development.} 
Proliferating chondrocytes $\mathrm{C}$, (shown in red) proliferate or differentiate into hypertrophic chondrocytes $\mathrm{H}$ (shown in blue).
$\mathrm{H}$ express the secreted protein Indian Hedgehog IHH (symbol $\mathrm{I}$), and $\mathrm{C}$ its receptor PTCH ($\mathrm{R}$).
PTCH and IHH form a complex (symbol $\mathrm{R}^{2}\mathrm{I}$) and trigger processes controlling differentiation ($\mathrm{C}\rightarrow\mathrm{H}$).
Panel (\textit{A}) and (\textit{B}) illustrate two different regulatory hypotheses: in 
(\textit{A}), the inhibitory effect of IHH signaling is directly affecting differentiation (\textit{The Directly Coupled Model}), whereas in 
(\textit{B}) the IHH signaling leads to production of a hypothetical paracrine factor $\mathrm{P}$ that prevents differentiation (\textit{The Indirectly Coupled Model}).
(\textit{C}) Deforming tissue growth. Proliferating chondrocytes $\mathrm{C}$ (shown in red) divide, which is modeled as a local mass source $\mathcal{S}_{\mathrm{prol}}$ (left path).
As a result of hypertrophic differentiation, the cells increase in volume and lead to a local mass source
$\mathcal{S}_{\mathrm{diff}}$ (right path). Both mechanisms induce a velocity field $\boldsymbol{u}$
in the fluid, whose internal boundary is passively advected.

\paragraph{Figure 3: The Prescribed Growth Model.}
(\textit{A}) The initial geometry of the chondrogenic tissue together with the growth restricting walls.
(\textit{B-E}) The local growth rate is spatially uniform and decreases reciprocally proportional to the area of the chondrogenic tissue, leading to linear elongation of the domain.
(\textit{B}) The concentration levels of $\mathrm{R}^{2}\mathrm{I}$ at the ends emerge at $t\approx 19.7$ and stay stable until $t\approx 23.1$. Then IHH signaling increases in the middle, where \textit{Ihh} expressing hypertrophic chondrocytes accumulate. Upon epiphyseal differentiation at time $t\approx35.2$, high IHH signaling reappears at the ends.
(\textit{C-D}) The concentrations of proliferative and hypertrophic chondrocytes are close to complementary.
(\textit{E}) The differentiation rate leads to high hypertrophic differentiation in the center of the domain at time $t\approx22.9$, and at the ends at time $t\approx 34.7$.
(\textit{F-J}) The local growth rate is spatially uniform and constant, leading to exponential elongation.
(\textit{F}) The complex concentration at $t=40$. The midline concentration is also depicted in (\textit{G}).
(\textit{G}) Additional modes appear at time $t\approx 34.1$ in the complex concentration.
(\textit{H-J}) The additional modes in the complex concentration are reflected in the cell concentrations (\textit{H-I}) and the differentiation rate (\textit{J}).
The initial values are uniformly zero for $\mathrm{R}^{2}\mathrm{I}$ and the differentiation rate,
uniformly $0.9$ for $\mathrm{C}$, and uniformly $0.1$ for $\mathrm{H}$ (c.f. Table \ref{tab:parameters}).

\paragraph{Figure 4: The Directly Coupled Model.}
(\textit{A-E}) and (\textit{F-J}) show the same simulations with noisy initial cell concentrations.
(\textit{B}) The domain grows exponentilly. The radial symmetry of IHH signaling is lost at $t\approx 22.8$.
(\textit{C-D}) The proliferative chondrocytes differentiate into hypertrophic chondrocytes at the beginning; the domain is subsequently expanded by proliferation.
(\textit{E}) Differentiation is high at $t=0$ and zero otherwise.
(\textit{G-J}) The results are similar for different noisy initial conditions, but the Turing system cannot deterministically pattern the chondrogenic domain.
The initial values are uniformly zero for $\mathrm{R}^{2}\mathrm{I}$ and the differentiation rate,
uniformly $0.9$ for $\mathrm{C}$, and uniformly $0.1$ for $\mathrm{H}$ (c.f. Table \ref{tab:parameters}).

\paragraph{Figure 5: The Indirectly Coupled Model.}
(\textit{A}) The complex accumulates at the ends between time $t\approx 17.7$ and $t\approx 24.9$. The IHH signaling upregulates significantly upon differentiation in the center at time $t\approx 24.9$ and at the ends at time $t\approx 37.7$.
(\textit{E}) Proliferating chondrocytes express \textit{Pthrp} at high levels of IHH signaling.
(\textit{C}) \textit{Pthrp} expression leads to local accumulation of PTHrP, which inhibits cell differentiation.
(\textit{D}) At locations where PTHrP concentration is low, proliferative chondrocytes (shown in (\textit{E})) differentiate into hypertrophic chondrocytes (shown in (\textit{F})). The increase of cell volume causes accelerated longitudinal growth. Although the conditions in the center always allow differentiation for $t>24.9$, the majority of proliferative chondrocytes differentiate in a short time span.
(\textit{G}) Hypertrophic chondrocytes express \textit{Ihh}, which leads to high levels of IHH signaling (cf. (\textit{A}))
(\textit{H}) \textit{Ptch} expression is positively upregulated at locations of high IHH signaling.
The initial values are uniformly zero for $\mathrm{R}^{2}\mathrm{I}$, $\mathrm{P}$ expression
and the differentiation rate,
uniformly $1000$ for $\mathrm{P}$,
uniformly $0.9$ for $\mathrm{C}$, uniformly $0.1$ for $\mathrm{H}$,
uniformly $\gamma \rho_{\mathrm{I}} \mathrm{H}_{0}$ and $\gamma \rho_{\mathrm{R}} \mathrm{C}_{0}$
for $\mathrm{I}$ and $\mathrm{R}$ expression, respectively
(c.f. Table \ref{tab:parameters}).

\paragraph{Figure 6: The Indirectly Coupled Model with a threshold on Ptch production.}
\textit{Ptch} is not expressed any more when the concentration of proliferating chondrocytes is negligible.
(\textit{A}) IHH signaling is only high in transition zones, although \textit {Ihh} is expressed at high levels in the center.
(\textit{B-G}) The introduced threshold does not significantly affect \textit{Pthrp} and \textit{Ihh} expression, PTHrP and cell differentiation.
(\textit{H}) \textit{Ptch} expression is only upregulated at locations of high IHH signaling (cf. (\textit{A}))
The initial values are uniformly zero for $\mathrm{R}^{2}\mathrm{I}$, $\mathrm{P}$ expression
and the differentiation rate,
uniformly $1000$ for $\mathrm{P}$,
uniformly $0.9$ for $\mathrm{C}$, uniformly $0.1$ for $\mathrm{H}$,
uniformly $\gamma \rho_{\mathrm{I}} \mathrm{H}_{0}$ and $\gamma \rho_{\mathrm{R}} \mathrm{C}_{0}$
for $\mathrm{I}$ and $\mathrm{R}$ expression, respectively
(c.f. Table \ref{tab:parameters}).

\paragraph{Figure 7: Mutants.}
Parts of the network of the indirectly coupled model were altered to assess the effect on the phenotype.
(\textit{A}) \textit{Ihh} expression is shut off.
The complex fails to form a pattern, and so the downstream processes.
The tissue grows due to it's intrinsic proliferation.
When the initial PTHrP concentration falls below the differentiation threshold $\theta_{i}$, the entire primordium differentiates instantaneously, leading to highly accelerated growth.
(\textit{B-C}) \textit{Pthrp} expression is shut off.
The complex pattern is formed normally at time $t\approx 17.7$, but the downstream \textit{Pthrp} expression pattern cannot form.
The initial PTHrP is diluted and degraded, leading to spatially homogeneous, instantaneous differentiation at time $t\approx 24.9$.
(\textit{D-F}) \textit{Pthrp} is overexpressed tenfold.
(\textit{D}) The complex pattern forms normally at time $t\approx 17.7$ and stays stable until the end of the observed period of time at $t=40$.
(\textit{E}) The \textit{Pthrp} expression is regulated by the complex, thus entopic.
(\textit{F}) The PTHrP concentration is high, leading to global inhibition of differentiation.
Therefore, no spatial organization of the cells can be observed.
The initial values are uniformly zero for $\mathrm{R}^{2}\mathrm{I}$ and $\mathrm{P}$ expression,
uniformly $1000$ for $\mathrm{P}$,
and uniformly $\gamma \rho_{\mathrm{R}} \mathrm{C}_{0}$
for $\mathrm{R}$ expression for all mutants
(c.f. Table \ref{tab:parameters}).

\paragraph{Figure 8: Local Parameter Sensitivity for the Indirectly Coupled Model.}
The set point parameters from Table \ref{tab:parameters} were changed as long as
the central-lateral organization still appeared. The values were normalized to the set point parameters, and the factor by which the basal parameter value can be changed is shown.
The values were changed with $1\%$ accuracy if less than $10\%$, and with $10\%$ accuracy if less than $100\%$, and with $100\%$ accuracy otherwise.
The volume ratio $\Phi$ and the differentiation rate $\delta$ are not shown since they only affect the differentiative growth rate, but not the emergence of the central-lateral organization.
The diffusion coefficients of PTHrP, $D_{\mathrm{P}}$, and the cells, $D_{\mathrm{cells}}$, as well as the initial concentration of PTHrP, $\mathrm{P}_{0}$, can be increased massively without affecting the qualitative emergence of the central-lateral organization.

\clearpage
\newpage
\section*{Figures}

\newpage
\clearpage
\section*{Tables}

\begin{table}[h]
  \centering
  \normalsize
  \begin{tabular}{| l | l |}
  \hline
  IHH       				& the protein Indian Hedgehog\\
  \textit{Ihh}				& the gene encoding Indian Hedgehog\\
  \textit{Ihh} expression	& the production of the protein IHH\\
  IHH signaling				& the signal PTCH1$^{2}$IHH\\
  PTHrP						& the protein parathyroid hormone-related protein\\
  \textit{Pthrp}			& the gene encoding PTHrP\\
  \textit{Pthrp} expression	& the production of the protein PTHrP\\
  PTCH1						& the receptor for IHH\\
  \textit{Ptch1}			& the gene encoding PTCH1\\
  \textit{Ptch1} expression	& the production of the receptor PTCH1\\
  \hline
  \end{tabular}
  \caption{\textbf{Terminology.}}
  \label{tab:teminology}
\end{table}

\begin{table}[h]
  \centering
  \normalsize

  \begin{tabular}{| l | l | l |}
  \hline
  $L_{\mathrm{domain}}$       				& 20                     & size of bounding box \\
  \hline \hline
  $\gamma$ 									& 100		& scaling factor \\
  \hline \hline
  $D_{\mathrm{I}}$ 							& 10		& \multirow{3}{*}{diffusion coefficients}\\
  $D_{\mathrm{P}}$ 							& 0.01		& \\
  $D_{\mathrm{cell}}$ 						& 0.001		& \\
  \hline \hline
  $\varphi$      							& 0.0005 &  proliferation rate \\
  $\delta$ 									& 0.05		& differentiation rate\\
  \hline \hline
  $\Phi$									& 4		& volume ratio $\mathrm{H}/\mathrm{C}$\\
  \hline \hline
  $\rho_{\mathrm{I}}$						& 10		& \multirow{2}{*}{production rates}\\
  $\rho_{\mathrm{R}}$						& 0.1		& \\
  \hline \hline
  $\delta_{\mathrm{P}}$						& 0.001		& decay rate\\
  \hline \hline
  $\theta_{\mathrm{P}}$ 					& 1 	& threshold for $\mathrm{P}$ production\\
  $\theta_{d}$								& 0.02 	& \multirow{2}{*}{threshold for differentiation}\\
  $\theta_{i}$ 								& 30	& \\
  \hline \hline
  $\mathrm{I}_{0}$ 							& 0 		& \multirow{5}{*}{uniform initial values} \\
  $\mathrm{R}_{0}$ 							& 1 		&  \\
  $\mathrm{P}_{0}$ 							& 1000 		&  \\
  $\mathrm{C}_{0}$ 							& 0.9 		& \\
  $\mathrm{H}_{0}$ 							& 0.1		& \\
  $L_{\mathrm{bone}}\left(t=0\right)$       & 1                     & initial bone domain length \\
  $W_{\mathrm{bone}}\left(t=0\right)$       & 2/5                     & initial bone domain width \\
  \hline
  \end{tabular}

\caption{\textbf{Parameter Values.} Index $d$ and $i$ refer to the \textit{directly} and \textit{indirectly} coupled model, respectively. All the other parameters are the same for all three models.}
\label{tab:parameters}
\end{table}

\end{document}